\documentclass[
    aps,
    prx,
    reprint,
    superscriptaddress,
    amsmath,
    amssymb
]{revtex4-2}

\usepackage{graphicx}
\usepackage{tikz}
\usetikzlibrary{arrows.meta, positioning, fit}

\usepackage{booktabs}
\usepackage{multirow}
\usepackage{makecell}

\usepackage{amsfonts}
\usepackage{bbm}
\usepackage{nicefrac}

\usepackage{microtype}
\usepackage{xcolor}

\usepackage{hyperref}
\usepackage{url}

\usepackage{wrapfig}

\begin{document}


\title{Worse Forces Can Lead to Better Structures: Trajectory-Level Fine-Tuning of Machine Learning Interatomic Potentials}


\author{Evan Dramko}
\affiliation{
Department of Computer Science,
Rice University,
Houston, Texas 77005, USA
}

\author{Yizhi Zhu}
\affiliation{
Department of Materials Science and Nanoengineering,
Rice University,
Houston, Texas 77005, USA
}
\affiliation{
Rice Advanced Materials Institute,
Rice University,
Houston, Texas 77005, USA
}

\author{Yifei Zhang}
\affiliation{
Department of Computer Science,
Rice University,
Houston, Texas 77005, USA
}

\author{Aleksander Krivokapic}
\affiliation{Faculty of Technical Sciences, 
University of Novi Sad, 
Novi Sad, Serbia
}

\author{Geoffroy Hautier}
\email{gh55@rice.edu}
\affiliation{
Department of Materials Science and Nanoengineering,
Rice University,
Houston, Texas 77005, USA
}
\affiliation{
Rice Advanced Materials Institute,
Rice University,
Houston, Texas 77005, USA
}

\author{Thomas Reps}
\affiliation{
Department of Computer Sciences,
University of Wisconsin--Madison,
Madison, Wisconsin 53706, USA
}

\author{Christopher Jermaine}
\affiliation{
Department of Computer Science,
Rice University,
Houston, Texas 77005, USA
}

\author{Anastasios Kyrillidis}
\affiliation{
Department of Computer Science,
Rice University,
Houston, Texas 77005, USA
}


\begin{abstract}
Machine learning interatomic potentials (MLIPs) are commonly trained and evaluated by their ability to reproduce first-principles energies and forces. Their practical value, however, is often determined by downstream processes such as structural relaxation, in which predictions are repeatedly fed back into the model. Much of the MLIP literature is motivated by the expectation that improving local force accuracy will produce correspondingly better downstream behavior. Although this expectation is natural in the limit of a perfectly reconstructed potential-energy surface, it may not hold in practice due to inherent model inaccuracies. Here, we demonstrate that force accuracy and relaxation accuracy can be systematically driven in opposite directions. Starting from pretrained MLIPs, we differentiate through complete relaxation trajectories and optimize the accuracy of the final structures. This controlled intervention produces models whose force-prediction errors increase substantially while their relaxed structures become more accurate. These improvements cannot be reproduced by optimizing the relaxation step size alone. Importantly, this represents a trade-off between two aspects of physical fidelity: the learned forces become less faithful to first-principles forces, while the resulting relaxed structures become more faithful to first-principles predictions. Rather than merely observing a discrepancy between local prediction quality and downstream performance, we demonstrate that this discrepancy can be systematically induced through optimization. Our findings suggest that MLIPs need not be optimized exclusively as approximations of first-principles forces, but can instead learn task-specific vector fields that guide relaxation trajectories toward more accurate final structures. By propagating supervision through complete relaxation trajectories, our framework captures dependencies between successive predictions that are absent when structure--force pairs are treated independently. More broadly, these results demonstrate that conventional force metrics can provide an incomplete assessment of downstream performance and highlight the importance of evaluating scientific machine-learning models within the computational workflows in which they are deployed.
\end{abstract}


\maketitle

\section{Introduction}
Determining stable atomic configurations is a central task in computational materials science and chemistry. For fixed atomic species, structural relaxation seeks a local minimum of the potential energy surface (PES) by iteratively updating atomic coordinates according to the local energy landscape. The gradients of the PES correspond to interatomic forces, and first-principles methods such as density functional theory (DFT) provide a standard route for evaluating these forces with quantum-mechanical fidelity. The resulting calculations are expensive, however, and a full relaxation may require tens to hundreds of force evaluations. This cost becomes a major bottleneck in high-throughput workflows, particularly when large numbers of candidate structures must be screened or when repeated relaxations are embedded within broader materials-design pipelines \cite{huang2023central,hormann2025machine,xiong2026finding}.

Machine learning interatomic potentials (MLIPs), also referred to as machine learning force fields, address this bottleneck by approximating first-principles energies and forces at substantially lower computational cost \cite{batatia2022mace,chen2022universal,choudhary2021atomistic,deng2023chgnet,yang2024mattersim,cheon2020crystal,musaelian2023learning}. Since the Behler--Parrinello formulation \cite{behler2007generalized}, a large body of work has focused on increasingly expressive architectures and on incorporating physically motivated symmetries such as permutation, translation, and rotation invariance or equivariance. Modern approaches include message-passing and equivariant graph neural networks \cite{batatia2022mace,chen2022universal,deng2023chgnet,rhodes2025orb,batzner20223,fu2025learning,liao2022equiformer}, attention-based models \cite{liao2022equiformer,liao2023equiformerv2,liao2026equiformerv3,qu2026recipe,dramko2025adapt,jiao2026equivariant}, and simpler residual or coordinate-based networks that remain competitive in selected regimes \cite{behler2016perspective,artrith2016implementation,tam2019interatomic,dickel2021lammps,lopez2023aenet,kalayan2024neural,dramko2025adapt}. There is also growing interest in relaxing strict architectural constraints and allowing models to learn useful physical structure directly from data \cite{dramko2025adapt,elhag2025learning,kreiman2025transformers,abramson2024accurate,qu2024importance,hu2021forcenet}. In parallel, universal MLIPs aim to transfer across broad regions of chemical space \cite{yang2024mattersim,unke2021machine,batatia2022mace,chen2022universal}; nevertheless, domain-specific training and fine-tuning remain common when high accuracy is required for a particular materials class or downstream task \cite{baghishov2026application,kempen2025accurate,radova2025fine,deng2025systematic,deng2024overcoming,fu2022forces,ren2025research,jiang2024machine,torres2026machine,stippell2024building,sun2025prediction,dramko2025adapt}.

\begin{figure*}[t]
\centering

\begin{tikzpicture}[
    >=Stealth,
    node distance=1.2cm,
    mlip/.style={
        rounded corners=6pt,
        draw=teal!60!black,
        fill=teal!35,
        fill opacity=0.5,
        text opacity=1,
        minimum width=1.3cm,
        minimum height=1.0cm,
        font=\bfseries
    },
    arr/.style={->, line width=0.8pt, gray!70},
    trainbox/.style={
        rounded corners=8pt,
        draw=black,
        dashed,
        inner sep=6pt
    }
]

\node (cx0) {\includegraphics[width=2cm]{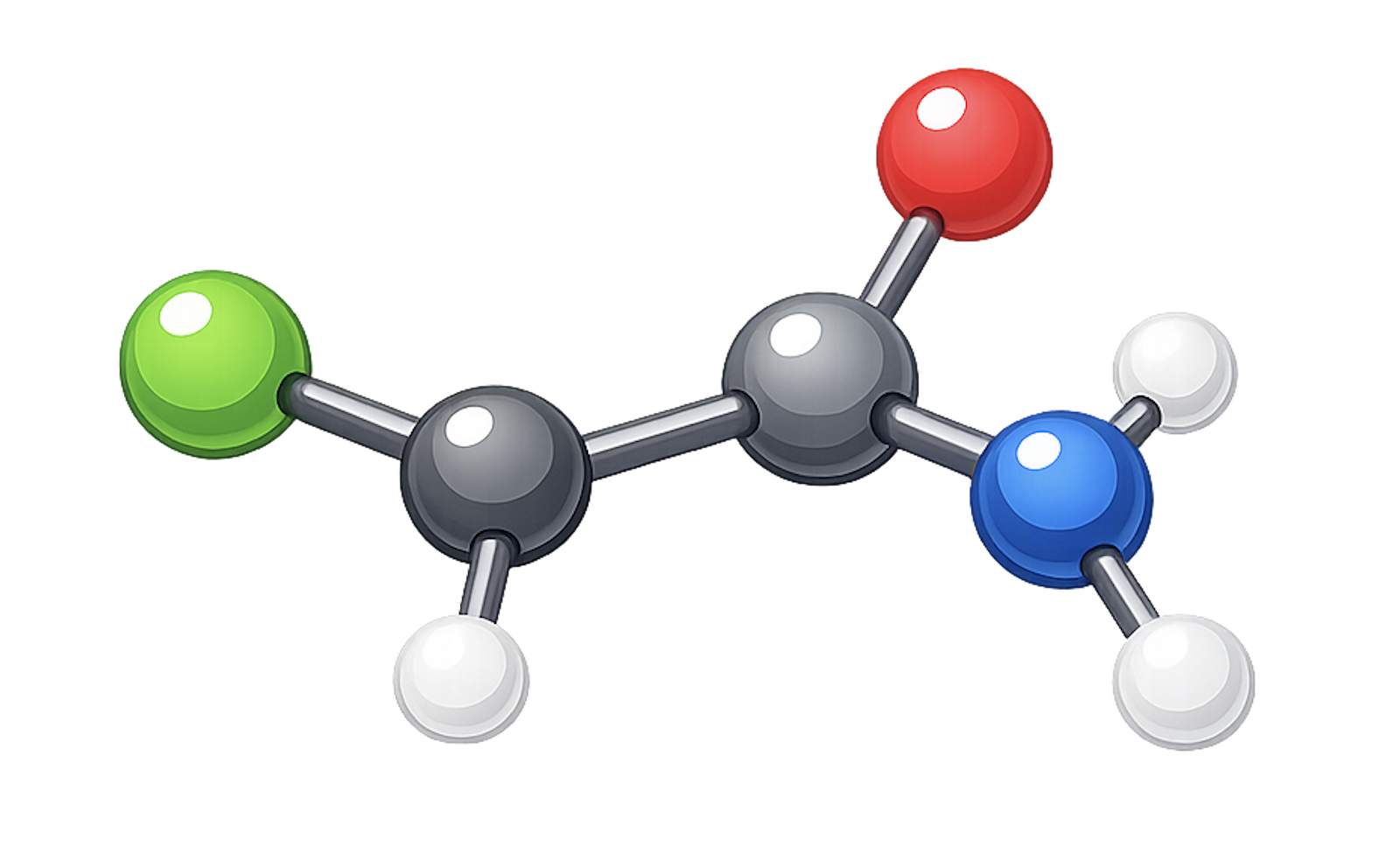}};
\node[mlip, right=of cx0] (cm1) {MLIP};
\node[mlip, right=of cm1] (cm2) {MLIP};
\node[right=0.6cm of cm2] (cdots) {\(\cdots\)};
\node[mlip, right=0.6cm of cdots] (cmT) {MLIP};
\node[right=of cmT] (cxT) {\includegraphics[width=2cm]{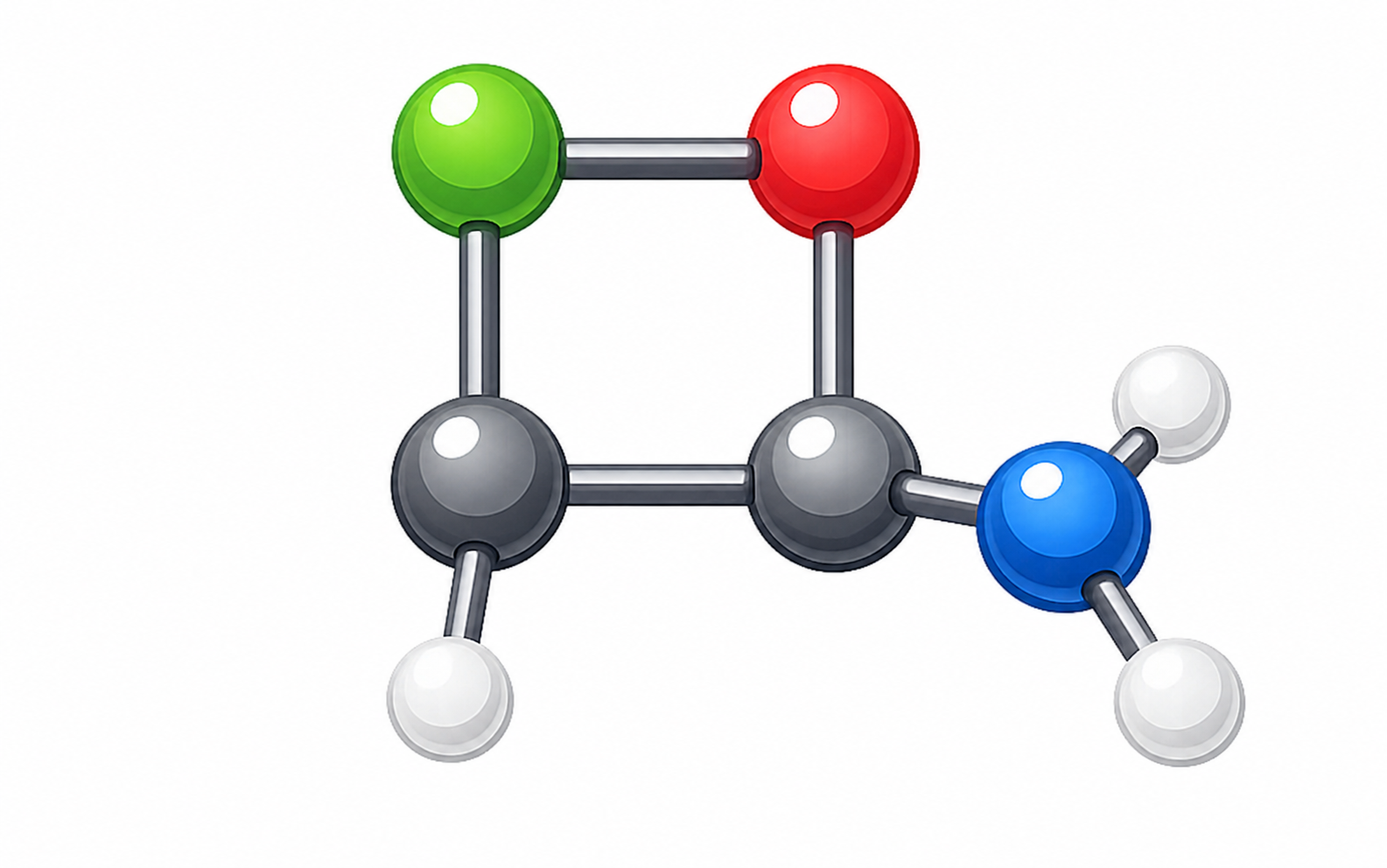}};

\draw[arr] (cx0) -- (cm1);
\draw[arr] (cm1) -- (cm2);
\draw[arr] (cm2) -- (cdots);
\draw[arr] (cdots) -- (cmT);
\draw[arr] (cmT) -- (cxT);

\node[below=0.35cm of cm1, font=\scriptsize, align=center] 
    {\(\mathbf{F}_0=f_\theta(\mathbf{x}_0)\)\\[-1pt]
     \(\mathbf{x}_1=\mathbf{x}_0+\eta\mathbf{F}_0\)};

\node[below=0.35cm of cm2, font=\scriptsize, align=center] 
    {\(\mathbf{F}_1=f_\theta(\mathbf{x}_1)\)\\[-1pt]
     \(\mathbf{x}_2=\mathbf{x}_1+\eta\mathbf{F}_1\)};

\node[below=0.35cm of cmT, font=\scriptsize, align=center] 
    {\(\mathbf{F}_{T-1}=f_\theta(\mathbf{x}_{T-1})\)\\[-1pt]
     \(\mathbf{x}_T=\mathbf{x}_{T-1}+\eta\mathbf{F}_{T-1}\)};

\node[
    trainbox,
    fit=(cx0)(cm1)(cm2)(cdots)(cmT)(cxT),
] {};

\draw[<-, blue!60, thick]
([yshift=0.8cm]cm1.west) --
([yshift=0.8cm]cm1.east)
node[midway, above, yshift=4pt, font=\normalsize] {train one step};

\draw[<-, blue!60, thick]
([yshift=0.8cm]cm2.west) --
([yshift=0.8cm]cm2.east)
node[midway, above, yshift=4pt, font=\normalsize] {train one step};

\draw[<-, blue!60, thick]
([yshift=0.8cm]cmT.west) --
([yshift=0.8cm]cmT.east)
node[midway, above, yshift=4pt, font=\normalsize] {train one step};

\node[anchor=west, font=\bfseries] at ([xshift=-0.3cm,yshift=1.05cm]cx0.north west) {(a) Classic one-step training};

\end{tikzpicture}

\vspace{0.55cm}

\begin{tikzpicture}[
    >=Stealth,
    node distance=1.2cm,
    mlip/.style={
        rounded corners=6pt,
        draw=teal!60!black,
        fill=teal!35,
        fill opacity=0.5,
        text opacity=1,
        minimum width=1.3cm,
        minimum height=1.0cm,
        font=\bfseries
    },
    arr/.style={->, line width=0.8pt, gray!70},
    trainbox/.style={
        rounded corners=8pt,
        draw=black,
        dashed,
        inner sep=6pt
    }
]

\node (bx0) {\includegraphics[width=2cm]{images/molecule.png}};
\node[mlip, right=of bx0] (bm1) {MLIP};
\node[mlip, right=of bm1] (bm2) {MLIP};
\node[right=0.6cm of bm2] (bdots) {\(\cdots\)};
\node[mlip, right=0.6cm of bdots] (bmT) {MLIP};
\node[right=of bmT] (bxT) {\includegraphics[width=2cm]{images/molecule2.png}};

\draw[arr] (bx0) -- (bm1);
\draw[arr] (bm1) -- (bm2);
\draw[arr] (bm2) -- (bdots);
\draw[arr] (bdots) -- (bmT);
\draw[arr] (bmT) -- (bxT);

\node[below=0.35cm of bm1, font=\scriptsize, align=center] 
    {\(\mathbf{F}_0=f_\theta(\mathbf{x}_0)\)\\[-1pt]
     \(\mathbf{x}_1=\mathbf{x}_0+\eta\mathbf{F}_0\)};

\node[below=0.35cm of bm2, font=\scriptsize, align=center] 
    {\(\mathbf{F}_1=f_\theta(\mathbf{x}_1)\)\\[-1pt]
     \(\mathbf{x}_2=\mathbf{x}_1+\eta\mathbf{F}_1\)};

\node[below=0.35cm of bmT, font=\scriptsize, align=center] 
    {\(\mathbf{F}_{T-1}=f_\theta(\mathbf{x}_{T-1})\)\\[-1pt]
     \(\mathbf{x}_T=\mathbf{x}_{T-1}+\eta\mathbf{F}_{T-1}\)};

\node[
    trainbox,
    fit=(bx0)(bm1)(bm2)(bdots)(bmT)(bxT),
] {};

\draw[
    <-,
    blue!60,
    thick
]
([yshift=0.05cm]bx0.north) --
([yshift=0.05cm]bxT.north)
node[midway, above, yshift=4pt, font=\normalsize] {training signal across trajectory};

\node[anchor=west, font=\bfseries] at ([xshift=-0.3cm,yshift=1.05cm]bx0.north west) {(b) Trajectory-level supervision};

\end{tikzpicture}

\caption{Comparison between classic one-step MLIP training and BPTT fine-tuning. Classic training supervises each MLIP step independently, while BPTT treats the full relaxation trajectory as one differentiable computational graph.}
\label{fig:classicvsbptt}
\end{figure*}

In most of these settings, MLIPs are trained by treating each structure--energy/force pair as an independent supervised example. This approach is natural when the objective is to reproduce the underlying first-principles potential energy surface (PES) as faithfully as possible. However, minimizing local prediction error does not necessarily minimize error in downstream structural relaxation. Relaxation is inherently sequential: the forces predicted at one configuration determine the next configuration visited, thereby influencing all subsequent force evaluations. Local prediction errors may therefore accumulate, cancel, or redirect the trajectory toward a different basin of attraction. Conversely, a model with less accurate instantaneous force predictions may nevertheless guide the relaxation process toward a more accurate final structure. This distinction motivates the central question of this work: \textit{can trajectory-level supervision improve the accuracy of relaxed structures even at the expense of local force accuracy?}

Previous work has demonstrated that force prediction accuracy alone is insufficient to characterize the physical fidelity of MLIP-driven molecular dynamics simulations \cite{fu2022forces}. These findings highlight a potential disconnect between local force accuracy and downstream performance, but leave open a more fundamental question: can these objectives be systematically driven in opposite directions through optimization of the same model? We address this question by introducing a differentiable trajectory-level fine-tuning framework that directly optimizes the final relaxed structure rather than agreement with reference forces. Starting from pretrained MLIPs, we demonstrate that this approach consistently improves structural relaxation accuracy while simultaneously increasing force prediction error across multiple architectures and datasets. By deliberately inducing this trade-off within individual models, our results establish that improvements in downstream structural accuracy need not be accompanied by improvements in local force fidelity. More broadly, they suggest that MLIPs can be optimized for their role within downstream computational workflows rather than exclusively for their accuracy as approximations of first-principles forces.

\subsection{Relaxation as a downstream objective}
MLIPs are commonly inserted into iterative optimization procedures rather than used to predict a relaxed structure in a single step. In practical workflows, they may also serve as a pre-relaxation stage before a final DFT refinement, reducing the number of expensive first-principles iterations required \cite{rossignol2023tern,dramko2025adapt}. Direct structure-prediction (DSP) methods have also been explored \cite{chanussot2021open,kolluru2022open,yang2024scalable}, but the strongly non-convex nature of atomic relaxation makes one-shot prediction difficult: prior studies report sensitivity to dataset bias, distribution shift, and falsely stable predictions, and competitive benchmark scores do not always translate to reliable end-to-end behavior \cite{shu2018amortized,sercu2021neural,riebesell2023matbench,malik2026made,jakob2025universally}. These limitations provide additional motivation for retaining the iterative structure of relaxation while reconsidering how the force model inside that loop is trained. To confirm that DSP is not viable for this task, we include results in Appendix \ref{sec:is2rs} showing that our iterative methods outperform DSP.

Two levels of optimization are therefore relevant. At the model level, the MLIP parameters are adjusted during training; at the physical-configuration level, atomic coordinates are iteratively updated during structural relaxation. Conventional MLIP workflows largely separate these levels: the force model is first optimized against local energy and force labels, after which its predictions are supplied to a separately chosen relaxation algorithm. Here, we couple the two levels by treating the complete relaxation trajectory as a differentiable computation. This allows the final structural error to propagate through the sequence of coordinate updates and exposes the MLIP parameters to the accumulated consequences of their intermediate predictions.

We study this formulation primarily as a fine-tuning strategy rather than as a replacement for conventional MLIP pretraining. This sequencing serves both an experimental and a practical purpose. First, it provides a controlled comparison between local force supervision and trajectory-level structural supervision. The pretrained model has already been optimized for structure--force matching; subsequent fine-tuning changes only the training objective, allowing us to examine how optimizing the final relaxed structure alters a model initially trained for local force accuracy. The resulting changes in force and relaxation performance can therefore be attributed to the shift in supervision rather than to differences in architecture or initialization. Second, trajectory-level supervision is ineffective when used to train a randomly initialized model from scratch, as shown in Section~\ref{sec:finetine}. This result suggests that final-structure supervision alone does not provide a sufficiently constrained or stable route to learning a useful force field. Instead, conventional pretraining supplies a physically grounded initialization that trajectory-level optimization can subsequently reshape for the downstream relaxation task.

Starting from this pretrained force model, we unroll a differentiable structural relaxation and use backpropagation through time (BPTT) to update the MLIP according to the error in the final relaxed structure. The procedure uses structural targets already present in relaxation datasets and requires no additional DFT calculations. Its change in performance therefore comes not from introducing new first-principles data, but from using existing data at a different level of supervision: final structures and the sequential dependence among relaxation steps constrain the force-like predictions made along the generated trajectory. We additionally investigate a memory-efficient formulation based on deep-equilibrium methods, motivated by the observation that converged relaxations are fixed points of an iterative map \cite{almeida1988backpropagation, bai2019deep}. Full BPTT is more effective in our experiments, while the equilibrium analysis helps distinguish two components of trajectory-level learning: steering the trajectory toward an appropriate basin of attraction and refining the configuration after that basin has been selected.

\subsection{Distinction from molecular dynamics and related trajectory learning}
It is important to distinguish the present objective from molecular dynamics (MD). Differentiable trajectory-level training has previously been used to match dynamical observables in molecular and coarse-grained simulations \cite{greener2024differentiable,greener2025reversible,krueger2024differentiable}, and related work has differentiated through numerical integration when learning Hamiltonian or differential-equation models \cite{chen2018neural,cranmer2020lagrangian,greydanus2019hamiltonian,chen2019symplectic,zhong2019symplectic,david2023symplectic}. Those settings require the learned forces or dynamics to retain physical meaning throughout a time evolution. Our goal is different: we optimize an iterative map for structural relaxation toward a local minimum. The fine-tuned model is therefore intended for relaxation, not for MD, and we do not claim that the modified forces provide a more faithful representation of instantaneous physical forces. Indeed, the central empirical result of this paper is precisely the opposite: force-prediction accuracy can decrease substantially after fine-tuning even as relaxed-structure accuracy improves.

Our formulation is also distinct from work that learns an optimizer or directly learns coordinate-update dynamics. For example, differentiable unrolled optimization has been used for molecular conformation prediction \cite{guan2021energy}. Here the relaxation procedure is not itself the object being learned; instead, we differentiate through an existing relaxation process in order to modify the underlying MLIP. The closest conceptual connection is to work on iterative maps, weight tying, and deep equilibrium models \cite{almeida1988backpropagation,bai2019deep,bai2021stabilizing}, as well as differentiable proxy objectives such as DOGSS \cite{yoon2020dogss}. The present setting differs in both objective and formulation: rather than replacing the underlying physical model or optimization procedure with a learned iterative map, we use differentiation through structural relaxation to fine-tune the MLIP itself, allowing the quality of the final relaxed structure to directly supervise its intermediate force predictions.

The contributions of this work are twofold. First, we introduce a trajectory-level fine-tuning framework for pretrained MLIPs that directly optimizes structural relaxation using differentiable rollouts, with both fully unrolled BPTT and an implicit-equilibrium variant. We analyze the role of step-size parameterization, initialization, trajectory length, memory cost, fixed-point behavior, and model architecture, and demonstrate the approach across multiple datasets and MLIP families. This technique improves downstream relaxation results without requiring any addtitional DFT calculations, by using information that is normally discarded in MLIP training. Second, and more broadly, we show that structural accuracy and instantaneous force accuracy can be systematically decoupled: a deliberately fine-tuned model can produce less accurate forces yet better relaxed structures. This implies that we can train models to improve physical fidelity to one measure, while breaking its fidelity to a different metric. Even for practitioners who do not intend to use trajectory-level fine-tuning, this provides a concrete reason to examine how MLIPs are evaluated and to distinguish fidelity to local force labels from performance on the downstream relaxation task.

\section{Method}
\subsection{Preliminaries} \label{sec:prelim}
\textbf{Notation.} In a slight abuse of notation, we denote \(\mathbf{X}\) as the atomic structure under consideration, and also as an \(\mathbbm{R}^{n \times 3}\) matrix of their coordinates. The ground truth relaxed structure is given by \(\mathbf{X}^*_F\), and the predicted structure is \(\mathbf{\widehat{X}_F}\). We denote DFT-calculated ground-truth forces with \(\mathbf{y}\) and predicted forces as \(\mathbf{\widehat{y}}\).

\textbf{Evaluation Metric.} Many works use MSE as a measure for structural error \cite{riebesell2023matbench}. While we show in Appendix \ref{sec:rmseLoss} that our technique is applicable for MSE-based evaluations as well, we instead primarily use a mass-weighted RMSE metric, defined in literature as Delta Q (\texttt{Dq}). \texttt{Dq} emphasizes physically meaningful structural deviations and is often used to quantify structural shift in first-principles calculations \cite{alkauskas2014first, somjit2025nv, davidsson2024discovery, alkauskas2014first1, alkauskas2014first2, turiansky2021computing, ali2023high}. It is defined by
\begin{equation}
\texttt{Dq}(\mathbf{X^*_F}, \widehat{\mathbf{X}}_F)
= \sqrt{\sum_{j=1}^{n} m_j \|\mathbf{X}^*_{j,:} - \widehat{\mathbf{X}}_{j,:}\|_2^2},
\end{equation}
where \(m_j\) denotes atomic mass, and \(i\) indexes the coordinates. 

\subsection{Relaxation as an Iterative Map}
We define a single relaxation step (\texttt{frame}) as:
\begin{equation}
\texttt{frame}(\mathbf{X}_t) \rightarrow \mathbf{X}_{t+1} \overset{\text{def}}{=} 
\begin{cases}
\widehat{\mathbf{y}} = \mathrm{MLIP}(\mathbf{X}_t) \\
\mathbf{X}_{t+1} = \mathbf{X}_t + \eta_t \widehat{\mathbf{y}}
\end{cases}
\end{equation}

For step size \(\eta_t\). 

A full relaxation trajectory is obtained by composition:
\begin{equation}
\widehat{\mathbf{X}}_F = \texttt{frame}^k(\mathbf{X}_0),
\end{equation}
where \(k\) is determined by a convergence criterion (not differentiated through). In the extreme, relaxation corresponds to a fixed-point iteration:
\begin{equation}
\lim_{k \to \infty} \texttt{frame}^k(\mathbf{X}_0) = \mathbf{X}^*.
\end{equation}

\subsection{BPTT Fine-Tuning}

We train MLIPs by backpropagating through the entire relaxation trajectory, with loss function:
\begin{equation}
\mathcal{L} = \texttt{Dq}(\mathbf{X}^*_F, \widehat{\mathbf{X}}_F).
\end{equation}

This differs from standard MLIP training, which optimizes per-step force prediction. Instead, we directly optimize the final relaxed structure, treating the full trajectory as a differentiable computation graph.

\subsection{Step-Size Parameterization}
In practice, a wide variety of optimizers are used in relaxations, however FIRE \cite{bitzek2006structural} is among the most common choices and is considered state-of-the-art. It monitors the alignment between forces and velocities and adaptively adjusts the dynamics by updating the time step, preserving momentum during consistent downhill motion and damping it when the motion becomes unstable. Because FIRE is non-differentiable, it cannot be used in a end-to-end BPTT training loop. We include it to establish a baseline performance for comparison with BPTT techniques. For use in BPTT we consider:

\begin{itemize}
\item Fixed step size \(\eta\)
\item Learned scalar step size
\item Learned step size via neural network
\end{itemize}


Experimental results in Table \ref{tab:stepSize} show that using a scalar step size produces better results as compared to a network learning the step size. 

\begin{table}[t]
\caption{
Ablation studies with ADAPT on the Si defect dataset for determining PES-level step sizes.
}
\label{tab:stepSize}
\centering

\begin{tabular}{lc}
\toprule
\textbf{Type} & \textbf{Average \texttt{Dq}} \\
\midrule
No BPTT                & 5.32 \\
Learned Scalar         & 2.73 \\
Decoder (Single Value) & 3.37 \\
\bottomrule
\end{tabular}

\end{table}

\subsection{Memory-Efficient Implicit Trajectory Supervision} \label{sec:deq_sec}
A practical limitation of trajectory-level supervision is the memory cost of backpropagation through time (BPTT), which scales linearly with the length of the relaxation trajectory because every intermediate structure must be retained for gradient computation. Since structural relaxations are naturally formulated as fixed-point iterations, an appealing alternative is to replace the fully unrolled backward pass with implicit differentiation at the converged equilibrium. This follows the philosophy of Deep Equilibrium Models (DEQs), which compute gradients through the fixed-point condition rather than through every iteration.

We therefore investigated a hybrid formulation in which only the initial portion of the relaxation trajectory is differentiated with BPTT, while the remainder of the relaxation proceeds without storing intermediate states, computes gradients using the implicit function theorem at convergence. The motivation is that early iterations largely determine the basin of attraction reached by the relaxation, whereas later iterations primarily refine the solution within that basin. When this decomposition is valid, trajectory-level supervision can be obtained with substantially lower memory requirements.

Although this formulation significantly reduced memory consumption, it situationally underperformed full BPTT in experiments. We hypothesize that equilibrium gradients primarily improve convergence within an already selected basin, whereas the structural improvements observed throughout this work arise from modifying the trajectory itself. Consequently, implicit differentiation appears to capture only part of the supervision signal contained within relaxation trajectories. While this approach is an interesting direction for reducing memory requirements, our experiments suggest that full trajectory-level supervision remains the most effective way to exploit the additional information present in structural relaxation trajectories.

An in depth analysis of DEQs in this context is provided in Appendix \ref{sec:gdeq}. A key finding from it is that trajectory supervision consists of two distinct learning problems: steering the trajectory into the correct basin of attraction and refining the equilibrium once it arrives.

\section{Results} \label{sec:results}
\begin{figure*}
\centering
    \includegraphics[width=\linewidth]{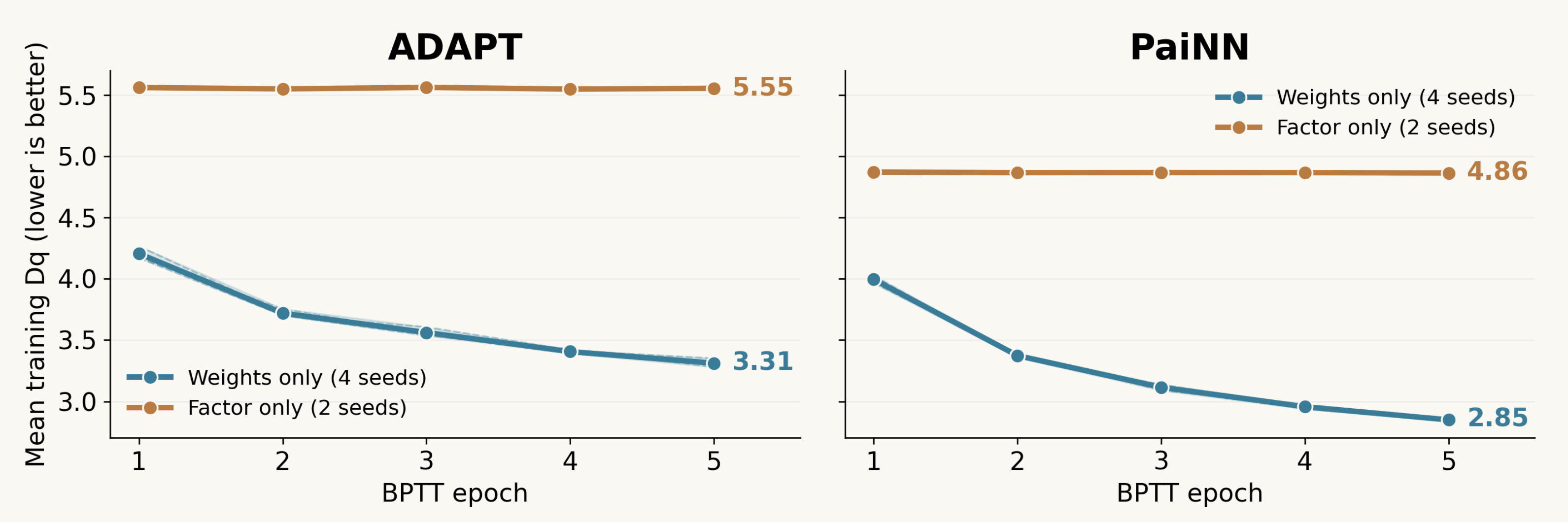}
    \caption{BPTT finetuning cannot match improvements by solely adjusting only the PES-level descent parameters.} \label{fig:learningSetupA}
\end{figure*}

\begin{figure*}
    \centering \includegraphics[width=0.9\linewidth]{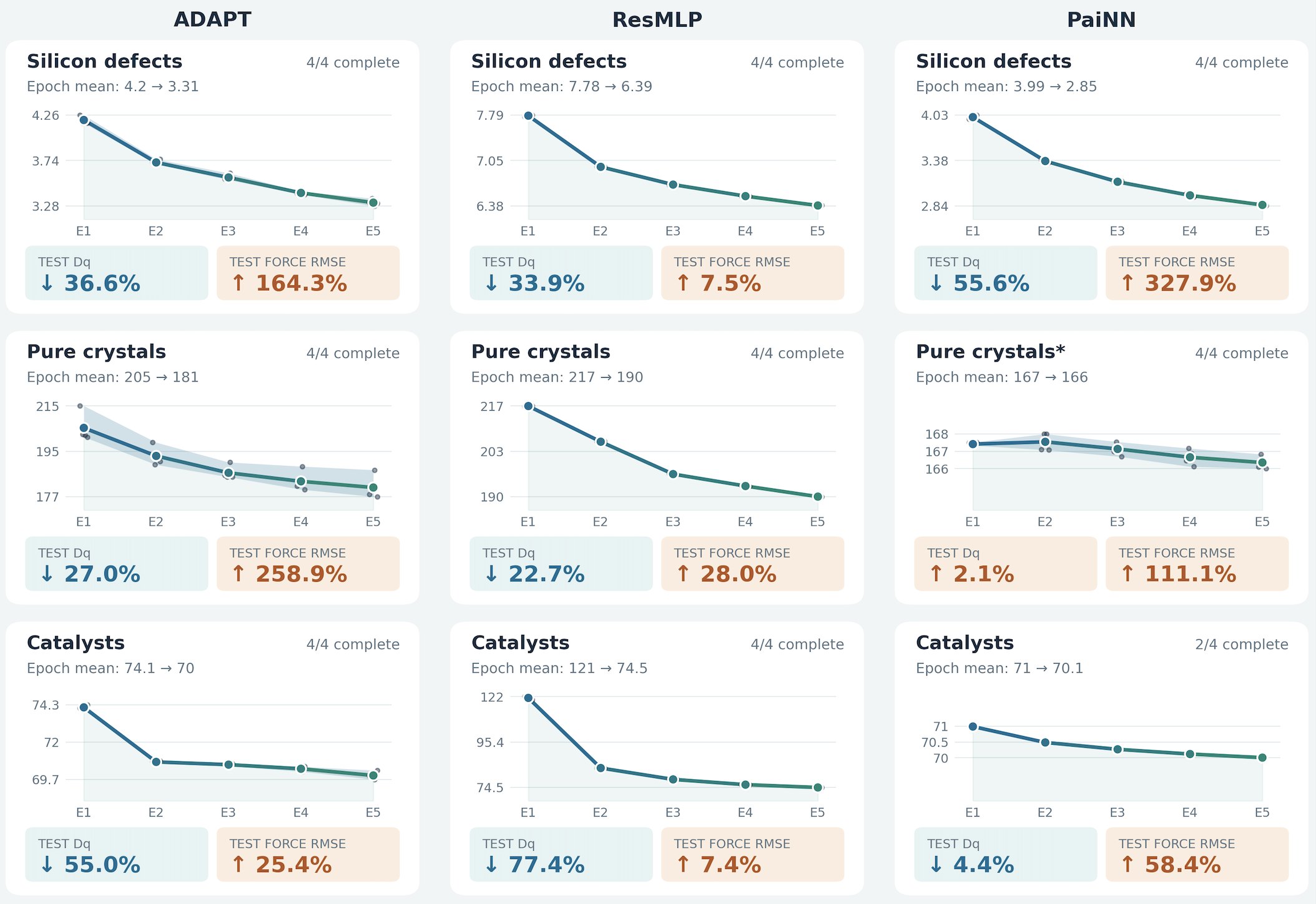} \caption{Test Dq structure error during training. Force RMSE is computed as the difference between the pre- and post-finetuning models on the test set. *In the PaiNN and Catalysts sample, the bottom 95\% of structures are used to handle extreme outliers. See Appendix \ref{sec:painnNotes} for more information.} \label{fig:trainingCurves}
\end{figure*}

The key comparison in this work is between a model before and after trajectory-level fine-tuning, which directly isolates the effect of the procedure. As a result, the appropriate baseline is the original, pre-tuned model, which allows us to measure the direct impact of trajectory-level fine-tuning independent of model architecture, size, or any other confounding factors.

We evaluate across three datasets: (1) crystal defects, (2) pure crystals and (3) catalysts. Collectively, these datasets encompass a wide range of applications and modeling scenarios, including structures spanning from fewer than 10 atoms to several hundred atoms, differing chemistries, and task-specific vs general purpose dataset. Details on the content and sourcing of the datasets are available in Appendix \ref{sec:datasets}. We note that there are few datasets viable for this study, as we require the dataset to contain full trajectories rather than individual frames. Datasets like MD22 \cite{chmiela2023accurate} which contain only a few final structures are intentionally excluded because they would provide an unfairly easy test case. To show robustness, we focus on datasets that exhibit substantial diversity in final structures. 

We consider three different model architectures: (1) ADAPT \cite{dramko2025adapt}, (2) ResMLP \ref{sec:MLIParchs}, and (3) PaiNN \cite{schutt2021equivariant}. Summaries of model architectures and justification of model relevance are available in Appendix \ref{sec:MLIParchs}. These cover a variety of techniques, including both graph and non-graph architectures, models used on both small and large scale applications, and collectively capture the key components of nearly all modern MLIPs, including equivariance constraints, message-passing, graph structure, attention computations, and residual connections. 
We adopt a fixed training horizon of five epochs for all experiments. Figure \ref{fig:trainingCurves} shows training loss exhibits rapid initial convergence followed by a pronounced reduction in its rate of improvement, consistent with entry into an asymptotic optimization regime. This choice also standardizes the training horizon across experimental conditions, supporting controlled comparisons while avoiding the disproportionate computational expense associated with long-tail optimization. Additional implementation details and reproducibility information are provided in Appendix \ref{sec:hyperparam}. 

\begin{figure*}
    \centering
        \includegraphics[width=\linewidth]{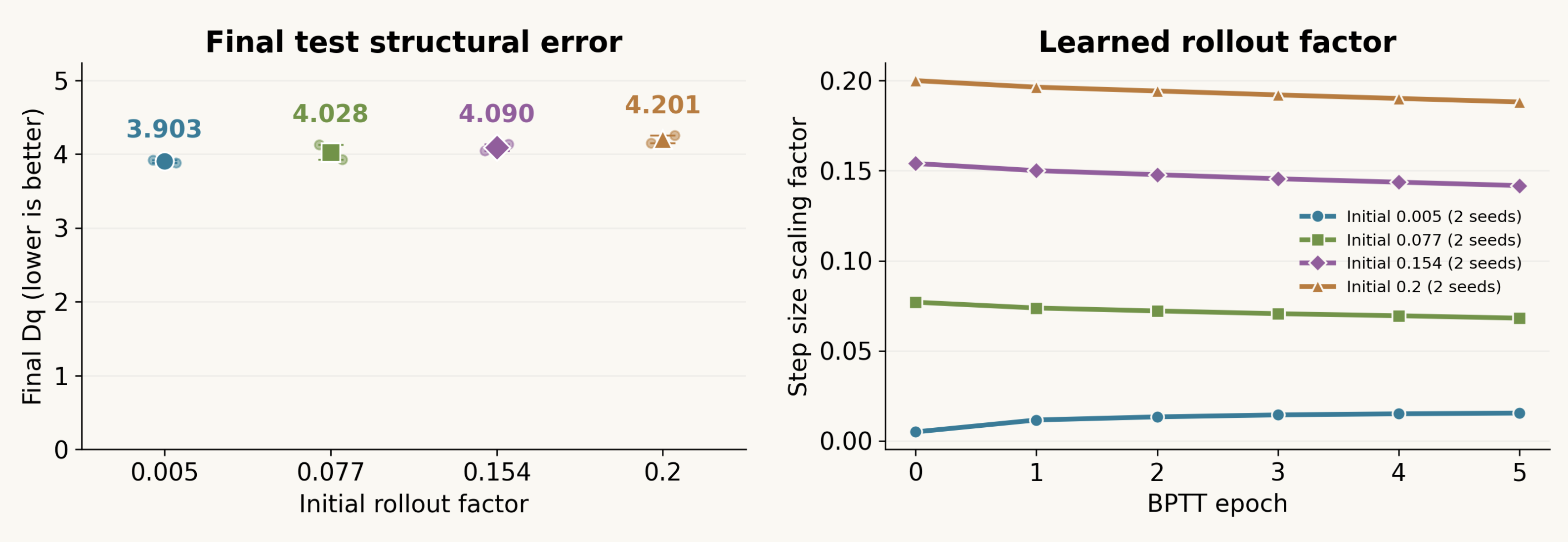} \caption{Effect of varying the learned PES-step parameter initializations and its evolution over the finetuning process. } \label{fig:initEffect}
\end{figure*}

\begin{figure}
    \centering
    \includegraphics[width=\linewidth]{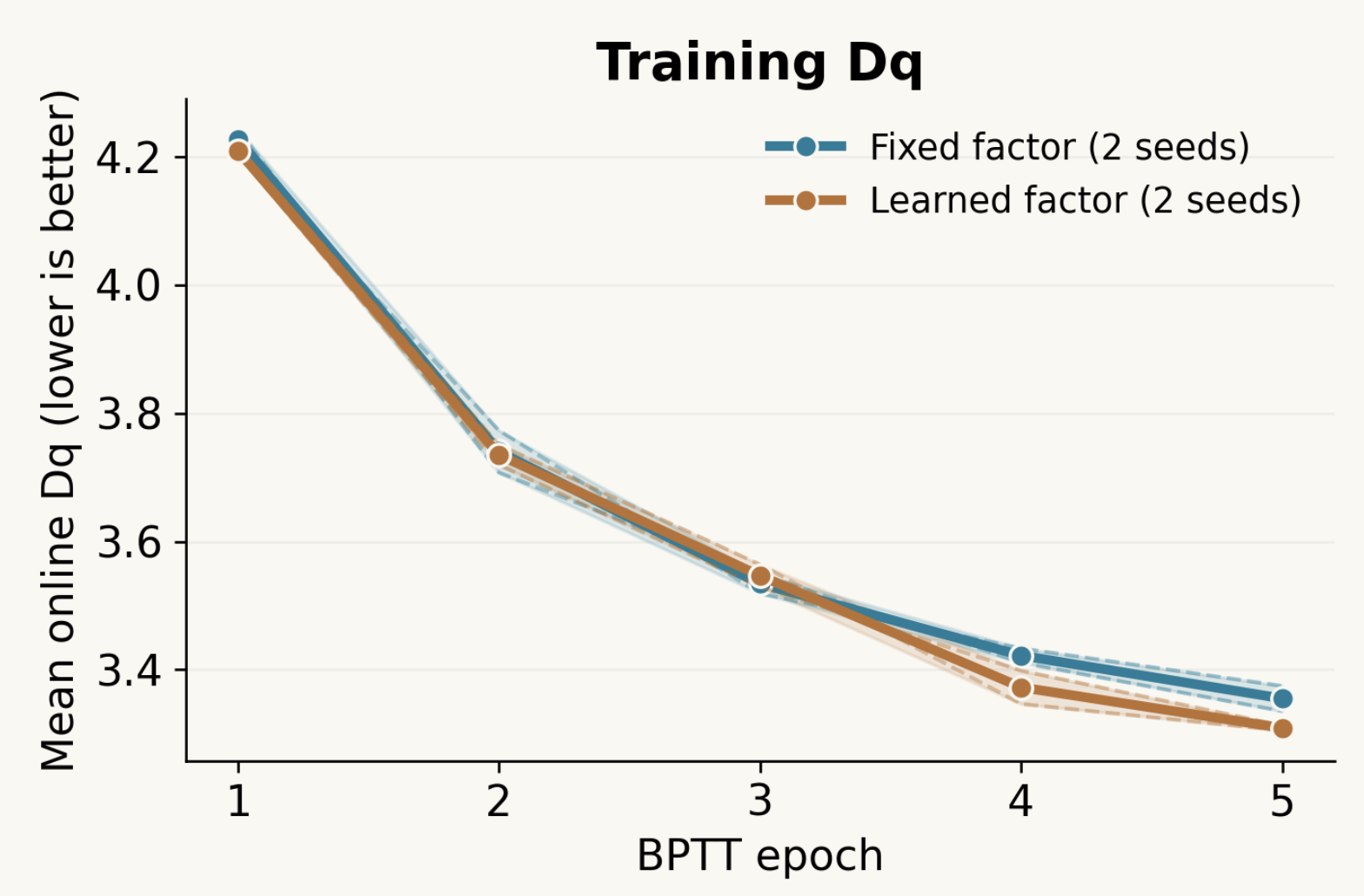}
    \caption{Experimental comparison of learned versus fixed scalar step sizes for ADAPT on Si defect relaxations. Error is reported as total Dq across the test set.}
    \label{fig:stepSearch}
\end{figure}

\subsection{Step Size Selection} \label{sec:stepSec}

\begin{figure*}
    \centering \includegraphics[width=\linewidth]{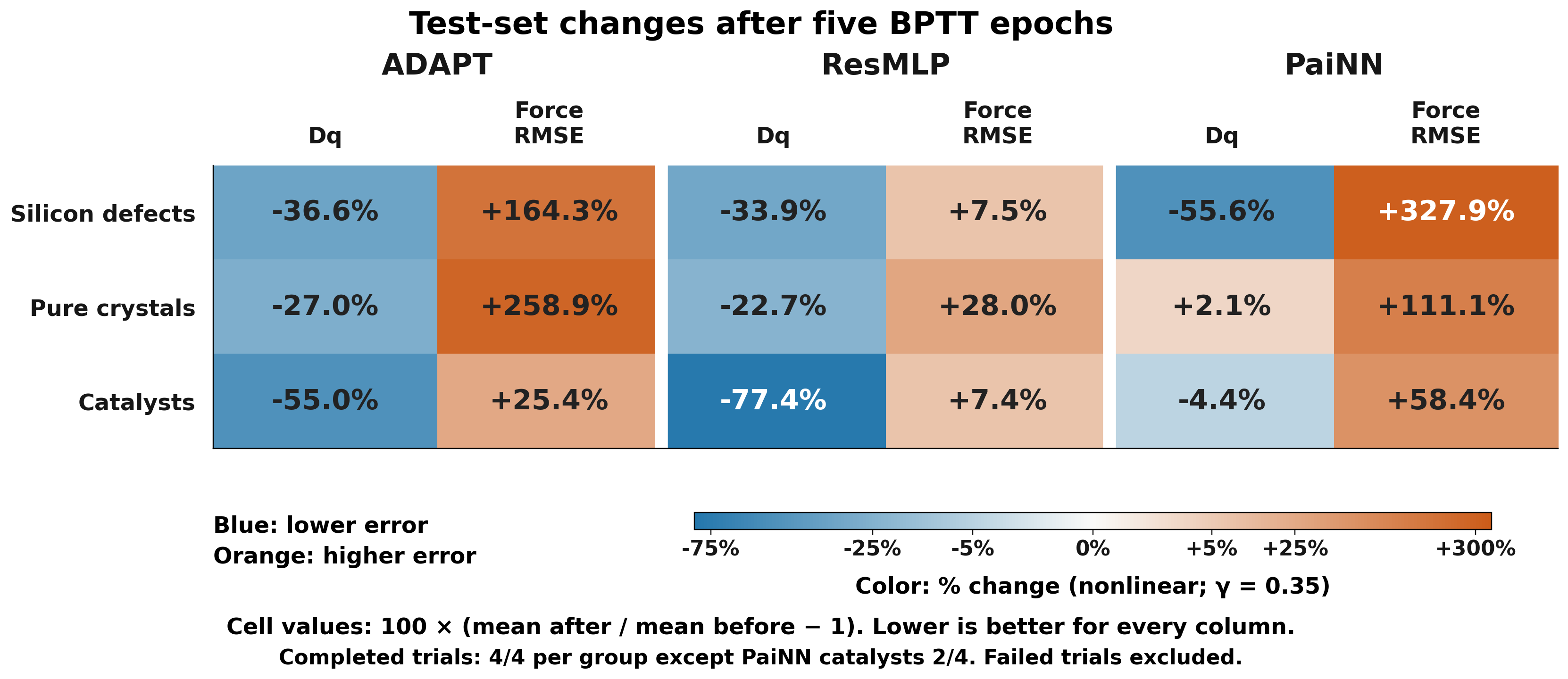} \caption{Effect of trajectory-level finetuning on structural relaxation and force prediction accuracy. See Appendix \ref{sec:painnNotes} for in detail analysis of PaiNN results.} \label{fig:heatmap}
\end{figure*}

We perform a grid search over step sizes on held-out evaluation data to estimate an upper-bound baseline for the PES-level step size (Appendix \ref{sec:PesStep}), finding that \(0.077\) yields the best results. This serves as an estimate of the base model’s strongest achievable performance without trajectory-level fine-tuning. Using this baseline, we show that BPTT fine-tuning is not simply adjusting hyperparameters, but instead creates substantive changes in the model’s predictions, a result confirmed by Section \ref{sec:forceStruct}, and Figure \ref{fig:heatmap}. As shown in Figure \ref{fig:initEffect}, BPTT is robust to the initialization of the PES-level step: even when initialized far from the optimal value, the model converges to comparable loss, requiring only a few more epochs of training time. We also show in Figure \ref{fig:stepSearch} that allowing for a fixed vs learned PES-level step size does not meaningfully impact model performance. In fact, even when exposed to gradients, the step size  changes very little, as seen in Figure \ref{fig:initEffect}. This indicates that the training ``prefers'' to update the MLIP weights to adjust for the PES-level step size, rather than modify the step size.\footnote{Notice that simple rescaling of the last weight matrix is equivalent to changing the PES-level step size.} Matching this, Figure \ref{fig:stepSearch} shows that keeping the step size fixed and not exposing it to gradient updates does not meaningfully impact performance. Figure \ref{fig:learningSetupA} further reinforces the cause of BPTT improvements. It shows that if only the PES-level descent parameter is exposed to gradient updates and not the model weights, then there is little to no meaningful performance gain. 

We also show in Appendix \ref{sec:ablation} that factors such as momentum, annealing the step size, and varying the trajectory length do not produce performance gains, and can even substantially hinder model performance.

\subsection{Result Interpretation}
\begin{figure}
    \centering
    \includegraphics[width=0.38\textwidth]{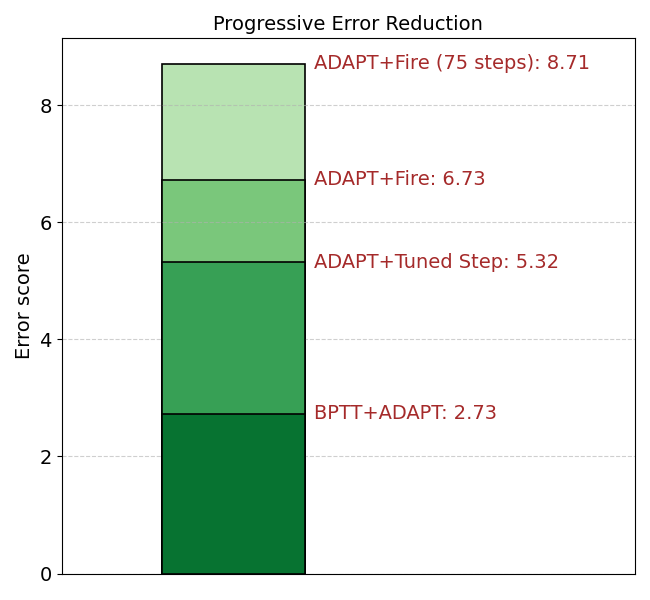}
    \caption{Effect of training schemes from sampled Si defect relaxations}
    \label{fig:improvementBar}
\end{figure}

Figure \ref{fig:heatmap} reveals a consistent trend across the evaluated models and datasets: trajectory-level fine-tuning improves the accuracy of relaxed structures while simultaneously degrading force prediction accuracy. This behavior is observed in the majority of the nine experimental settings. The principal exception is PaiNN on the Catalysts dataset, where a small number of outliers, potentially arising from numerical instabilities associated with repeated application of the model, substantially influence the aggregate results. Excluding these outliers reveals improved structural accuracy on the training set, although a modest degradation persists on the test set. Force prediction errors, meanwhile, increase substantially on both sets, matching expectations. Further analysis is provided in Appendix \ref{sec:painnNotes}.

These results demonstrate that the downstream performance of machine-learned interatomic potentials (MLIPs) need not be aligned with their local force prediction accuracy. Although accurate force predictions are generally desirable, minimizing force prediction error does not necessarily yield the best performance in downstream computational tasks. Fu et al. \cite{fu2022forces} previously demonstrated that force accuracy alone is insufficient to assess the physical fidelity of MLIP-driven molecular dynamics simulations, motivating evaluation through physical observables. While their work identifies this discrepancy through simulation-based evaluation, we introduce a generalizable optimization procedure that systematically exploits and increases the distinction between local force accuracy and downstream task performance.

Specifically, our fully differentiable trajectory-level fine-tuning framework propagates errors in the final relaxed structure through the entire relaxation trajectory, allowing the MLIP to be optimized directly for structural accuracy rather than agreement with reference forces. Across multiple architectures and datasets, this approach consistently improves relaxed structure predictions despite substantial degradation in force prediction accuracy. \textit{Importantly, this represents a trade-off between two aspects of physical fidelity: the learned forces become less faithful to the underlying first-principles forces, while the resulting relaxed structures become more faithful to first-principles structural predictions.} Rather than merely observing a discrepancy between local prediction quality and downstream performance, we demonstrate a systematic means of producing it through task-specific optimization. These findings suggest that MLIPs need not be optimized exclusively as approximations of first-principles forces, but can instead be tailored to their role within broader computational workflows, even when doing so sacrifices accuracy in intermediate physical predictions.

This perspective has practical implications for applications in which structural relaxation is the primary objective. For example, an MLIP intended to relax point defects in silicon may benefit more from optimization of final structural accuracy than from further reductions in force prediction error. Our method provides a general framework for such task-specific fine-tuning without requiring improvements in force fidelity. Furthermore, Figure \ref{fig:improvementBar} shows that, in the evaluated setting, our approach also achieves better structural relaxation accuracy than the widely used FIRE optimizer \cite{bitzek2006structural}.

\subsection{Qualitative Analysis of Crystal Defects} 
Of particular note in establishing technique validity is the study of ADAPT on crystal defects. Visual analysis of structures from the test set shows BPTT has a significant influence on the localization of interstitial defect atoms in the test cases. These observations confirm that the improvements are not attributable to trivial stabilization of the bulk lattice, which would be unlikely to produce a substantial reduction in the number of DFT steps required for full relaxation. It is shown in \cite{dramko2025adapt} that increasing the size and training time of the pretrained ADAPT model on this dataset does not improve its performance; thus establishing that BPTT fine-tuning is achieving accuracy in relaxations that would have been impossible with regular MLIP training.

\subsection{Physical Fidelity of tuned MLIPs} \label{sec:forceStruct}



BPTT shifts the objective of MLIP training away from local force accuracy toward trajectory-level correctness by learning an iterative update rule that preserves the fixed points and basins of attraction of the underlying PES. As a consequence, it introduces a clear trade-off: force prediction accuracy degrades after fine-tuning (Figure \ref{fig:heatmap}), while structural accuracy improves. This indicates that BPTT is not enhancing local physical fidelity, but instead capturing dataset-level structural trends and biasing predictions accordingly. The resulting behavior is well aligned with rapid structural relaxation tasks, but is unlikely to be suitable for molecular dynamics simulations, where accurate local force estimates are essential.


\subsection{Fine-tuning and Initialization Effect} \label{sec:finetine}
\begin{figure}
 \centering
        \includegraphics[width=\linewidth]{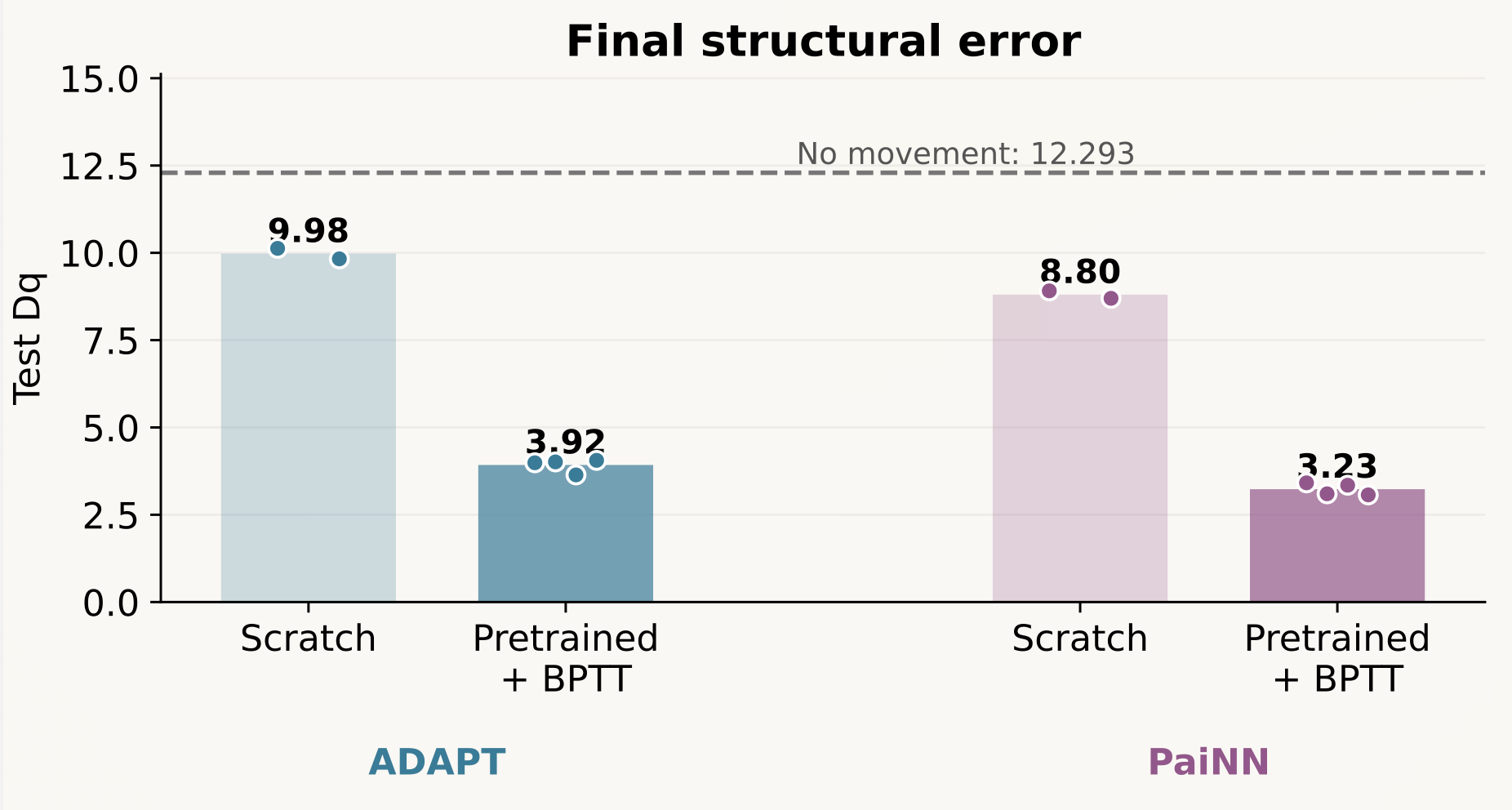}
        \caption{BPTT performance when training from random MLIP initialization compared with finetuning.} \label{fig:learningSetupB}
\end{figure}

Figure \ref{fig:learningSetupB} presents an experiment in which a randomly initialized model is trained using BPTT, without prior structure-to-force pretraining. The models trained from scratch for an equal number of epochs are substantially less capable than the pretrained models. This result is consistent with prior work on fixed-point learning (Section \ref{sec:deq_sec}), which highlights the importance of appropriate initialization for stable and effective training. From this perspective, a DFT-based relaxation procedure or a pretrained universal MLIP can be viewed as providing a reference fixed-point iteration, which our framework seeks to approximate through a task-specific model tailored to a restricted domain of chemical structures. 

Beyond demonstrating that local force accuracy and downstream relaxation performance can be systematically decoupled, our approach also offers a practical strategy for improving task-specific MLIPs. Despite advances in universal interatomic potentials, domain-specific models remain widely used in computational materials science \cite{baghishov2026application,kempen2025accurate,radova2025fine,deng2025systematic,deng2024overcoming,fu2022forces,ren2025research,jiang2024machine,torres2026machine,stippell2024building,sun2025prediction,dramko2025adapt}. Our framework provides a means of fine-tuning such models directly for structural relaxation, potentially improving their utility in specialized applications without requiring corresponding improvements in force prediction accuracy.


    



\subsection{Memory and Compute Usage} \label{sec:memory}
BPTT fine-tuning unrolls the network over \(t\) steps, increasing memory usage by a factor of \(t\) relative to a single force evaluation. Compute time likewise scales linearly with rollout length, with additional overhead from the PES-level step size and structure update operations. Despite this apparent cost, the overall training burden remains modest in practice. As shown in Figure \ref{fig:trainingCurves}, convergence typically occurs within three to five epochs, and Figure \ref{fig:stepSearch} shows that in some cases a single epoch suffices to capture the benefits. Moreover, fine-tuning operates only on initial and final states, rather than full trajectories, reducing the effective dataset size by approximately the average trajectory length.\footnote{Although not explored here, intermediate frames could be reused as new initial states to generate additional synthetic trajectories.}

In practice, forward pass batches are kept very small, even to a batch size of 1, to handle the memory requirements of storing the entire rollout. Training can be down with gradient accumulation to simulate larger batches. Thus, ignoring potential complicating factors from batch size selection, the total number of network evaluations scales as \(i \cdot j \cdot k\), where \(i\) is the number of epochs, \(j\) the average trajectory length, and \(k\) the number of trajectories. For the silicon defect setting, this yields \(5 \cdot 75 \cdot 6{,}000 \approx 2.25\)M forward passes through the MLIP. By comparison, standard training requires roughly \(250{,}000 \cdot 80 \approx 20\)M forward passes. Thus, although BPTT incurs higher per-step memory cost, its reduced epoch count and smaller effective dataset lead to lower overall compute than the classic MLIP training step.

Training long-trajectories is still feasible even if they exceed the memory capacity of the given hardware. Through the use of gradient checkpointing \cite{chen2016training}, training can be broken into blocks that fit into memory. 

If memory consumption becomes restrictive, Section \ref{sec:deq_sec} and Appendix \ref{sec:gdeq} show a feasible alternative approach using deep equilibrium models which have a linearly scaling memory consumption. The technique is less stable and ``turn-key'' than BPTT tuning however, leading us to favor BPTT in general.

\section{Discussion} \label{sec:discussion}
We present a method for fine-tuning MLIPs that uses structure information to modify force predictions. We calculate fully-differentiable relaxation trajectories using MLIPs, and then perform a BPTT update based on a structural accuracy metric. This process uses structure-to-structure supervision, which is a different aspect of the information than tradional structure-to-force prediction. 

The central result of this work is not simply that trajectory-level fine-tuning improves structural relaxation, but that relaxation accuracy can improve substantially even as conventional force accuracy becomes worse. In the silicon-defect setting, BPTT fine-tuning increases the force error from \(4.32\) to \(13.36\), while reducing the relaxation error from \(5.25\) to \(2.73\). Because this comparison is made before and after fine-tuning the same pretrained model, it isolates the effect of trajectory-level supervision from the architectural and training differences that complicate comparisons between independently developed MLIPs. Furthermore, these improvements persist on test trajectories excluded in their entirety from training, indicating that trajectory-level supervision can extract generalizable information rather than simply memorizing individual relaxation paths. These results demonstrate that downstream performance can be substantially improved with a fixed dataset and model architecture by extracting an additional level of supervision from existing structural information, rather than relying on more training data or increased model capacity.

Ablation studies further demonstrate that these improvements cannot be attributed simply to optimization of the relaxation procedure. Optimizing the step size without updating the MLIP does not reproduce the improvement, while updating the MLIP with a fixed step size produces substantial gains. Even a hand-selected step size chosen with knowledge of the evaluation set remains inferior to trajectory-level fine-tuning. Across the tested gradient-descent schemes, the model is capable of adapting its predicted forces to compensate for a fixed scalar step size, matching or outperforming more elaborate relaxation procedures. These results indicate that the improvement arises primarily from changes to the MLIP's predicted vector field, rather than changes to the relaxation optimizer itself.

This observation highlights a fundamental distinction between two possible roles of an MLIP: approximating the physical force field and providing effective update directions within a specified computational workflow. Conventional force-based pretraining emphasizes the former, whereas trajectory-level optimization can shift the model toward the latter. Mathematically, this distinction arises because multiple vector fields may transport the same initial structures toward similar endpoints, even when only one closely approximates the first-principles forces. Trajectory-level optimization can exploit these degrees of freedom, introducing deviations from the reference forces that would be penalized by conventional force matching but may be beneficial to the final structural objective. Consequently, an MLIP optimized for structural relaxation need not become a more accurate approximation of the underlying potential-energy surface. Instead, it may learn task-specific update directions whose repeated application more effectively guides structures toward their target configurations.

These findings have important implications for the training and evaluation of MLIPs. Force and energy accuracy remain essential for applications such as molecular dynamics, vibrational analysis, thermodynamic calculations, and transfer to configurations outside the targeted relaxation domain. However, local prediction accuracy alone is not necessarily sufficient to characterize performance in iterative downstream workflows. When model predictions are repeatedly composed within an optimizer or simulator, their accumulated effects may differ substantially from what single-step error metrics suggest. Benchmarks intended to measure relaxation performance should therefore supplement conventional force and energy errors with end-to-end metrics such as final-structure accuracy, basin recovery, convergence behavior, and the number of subsequent first-principles steps required to reach the reference minimum. More generally, the same dataset and model architecture can yield models optimized for substantially different notions of predictive utility.

Importantly, trajectory-level supervision is not proposed as a replacement for conventional MLIP training, but as a subsequent fine-tuning stage. It is particularly effective when applied to pretrained models in domain-specific settings, where accurate relaxation behavior over a restricted region of configuration space may be sufficient for the intended application. This makes trajectory-level fine-tuning particularly attractive in data-constrained settings, where generating additional first-principles calculations is computationally expensive. 

While trajectory-level supervision successfully decouples force accuracy from structural accuracy across multiple architectures and datasets, differentiating through long relaxation trajectories introduces substantial computational and numerical challenges. Long temporal rollouts create extreme effective network depth, and architectures without explicit mechanisms for stabilizing unrolled differentiation may exhibit unstable gradients. In our experiments, standard gradient clipping was insufficient to resolve these issues for PaiNN on the catalysts dataset, representing the only observed training failure. A detailed analysis of this case is provided in Appendix \ref{sec:painnNotes}. Developing equivariant GNN architectures that remain stable under long differentiable rollouts is therefore an important direction for future work, although it is distinct from the central finding that force accuracy and downstream relaxation accuracy can be systematically decoupled.

More broadly, these results demonstrate that accurate local force predictions and effective structural relaxation are distinct optimization objectives that need not improve together. By directly supervising the outcome of iterative relaxation, trajectory-level fine-tuning provides a means of improving downstream structural accuracy without additional training data or model capacity. This distinction motivates evaluating and optimizing MLIPs not only as approximations of physical force fields, but also in the context of the computational workflows in which they are deployed.

\section{Acknowledgments}
AI Usage: The authors used AI assistance in the review of existing literature, polishing and stylistic editing of written text, and to improve the style formatting of the plots. Core code and scripts were written by hand, but were reviewed and edited with AI, specifically codex sol, luna, and astra. The images of atomistic structures in Figure \ref{fig:classicvsbptt} were generated with AI; the rest of the diagram was made with TikZ. 

This research used resources of the National Energy Research Scientific Computing Center (NERSC), a Department of Energy User Facility using NERSC award AI4Sci@NERSC ERCAP0042057.

\newpage
\appendix
\section{ADAPT and ResMLP Architectures} \label{sec:MLIParchs}
\subsection{ADAPT}
We adopt the ADAPT architecture \cite{dramko2025adapt} as a pretrained MLIP test case due to both its effectiveness and its algorithmic simplicity. 
The model is defined as three-layer embedding MLP, a series of \(n\) Transformer encoder blocks and a final layer linear projection. Each encoder block (denoted \(\texttt{enc}\)) is given by: 
\begin{align}
\texttt{enc}(\mathbf{X}) \rightarrow \mathbf{X}_{out} = 
\begin{cases}
    \mathbf{H}_1 &= \texttt{LN}\bigl(\mathbf{X}_{\text{in}} + \texttt{Attn}(\mathbf{X}_{\text{in}})\bigr), \\
    \mathbf{H}_2 &= \texttt{FFN}\bigl(\texttt{LN}(\mathbf{H}_1)\bigr), \\
    \mathbf{X}_{\text{out}} &= \texttt{LN}(\mathbf{H}_2 + \mathbf{H}_1)
\end{cases}
\end{align}
ADAPT also uses projection and embedding operations defined by: 
\begin{align}
	\texttt{proj} &= \mathbf{W}_p\mathbf{X}\\
	\texttt{emb} &= \mathbf{W}_2 \circ \sigma \circ \mathbf{W}_1 \circ \sigma \circ \mathbf{W}_0 (\mathbf{X})
\end{align}
where \(\mathbf{W}_p \in \mathbbm{R}^{3 \times d_{model}}\), \(\texttt{emb}\) is a multi-layer perceptron (MLP), and \(\circ\) is the function composition operator. 
Note that the embedding does not use a positional encoding because the coordinates of the each atom is included in the atom features given to the model. The relative ordering of atoms does not matter. 

The overall ADAPT architecture is given by: 
\begin{align}
	\texttt{MLIP} &= \texttt{proj} \circ \underbrace{\texttt{enc} \circ ... \circ \texttt{enc}}_{\times n} \circ ~\texttt{emb} (\mathbf{X})
\end{align} 

\subsection{ResMLP} \label{sec:ResMLP}
Variations on the multilayer-perceptron architecture are a mainstay of MLIP architectures; having been used historically \cite{behler2007generalized, behler2016perspective, artrith2016implementation} and currently \cite{tam2019interatomic, dickel2021lammps, lopez2023aenet, kalayan2024neural, dramko2025adapt} to model the dynamics of atomistic calculations. We adopt a residual+MLP architecture as a testing scenario model, and denote it as ResMLP. 
\begin{align} \label{eq:resid}
    \mathbf{t}_0 &= \sigma(\mathbf{W}_{0} \mathbf{x} + \mathbf{b}_0)\\ \notag
    \mathbf{h}_0 &= \texttt{LN}(\mathbf{P}_{0} \mathbf{t}_0 + \mathbf{t}_0)\\ \notag
    \mathbf{t}_1 &= \sigma(\mathbf{W}_{1} \mathbf{h}_0 + \mathbf{b}_1)\\ \notag
    \mathbf{h}_1 &= \texttt{LN}(\mathbf{P}_{1} \mathbf{t}_1 + \mathbf{t}_1)\\ \notag
    &~~\vdots \\ \notag
    \mathbf{t}_5 &= \sigma(\mathbf{W}_{5} \mathbf{h}_4 + \mathbf{b}_5)\\ \notag
    \mathbf{h}_5 &= \texttt{LN}(\mathbf{P}_{5} \mathbf{t}_5 + \mathbf{t}_5)\\ \notag
    \mathbf{y} &= \mathbf{W}_6 \mathbf{h}_5 \notag
\end{align}

\noindent where each layer has \(4096\) nodes, and the output project down to \(220 \times 3 = 660\) values. While this may not be the most effective possible architecture in literature, variations of it are commonly used and it provides a strong argument for the generalizability of the BPTT fine-tuning results. 

\subsection{PaiNN}
We use an implementation of the PaiNN architecture based on the paper \cite{schutt2021equivariant}. Source code is attached in the released repository. This architecture was chosen because it features equivariant graph construction, a common theme in much of the modern MLIP literature. There is no specific property of PaiNN that motivated its selection as a representative GNN based MLIP. It has seen both historical and current use in practice \cite{stark2023machine, pelletier2026autotuning}.

\section{Scalar Decoder Head For Step Size} \label{sec:DecoderArch}
ADAPT provides the forces (gradient) of the PES (objective function), but does not tell us how large of a step we should take. We present a method that takes both the structure and forces as input, and outputs a step size for this iteration of the descent. This step-scaling and update procedure, \(\texttt{step}\) is defined in two parts: an upscaling projection to map from the native to embedding dimensions, and a scaler decoder head to produce a single step size for the whole structure.
The scaler decoder head, \(\texttt{dec}\) is defined by:
\begin{align}
\texttt{dec}(\mathbf{X}, \mathbf{q}) \rightarrow \mathbf{\widehat{y}} = 
\begin{cases}
    \mathbf{M} = \texttt{MLIP}(\mathbf{X}), \\
    \mathbf{h}_0 = \texttt{LN}(\mathbf{q} + \texttt{Attn}(\mathbf{q}, \mathbf{M}, \mathbf{M})), \\
    \mathbf{h}_1 = \texttt{LN}(\mathbf{h}_0 + \texttt{MLP}(\mathbf{h}_0)), \\
    \widehat{\mathbf{y}} = \mathbf{W}\mathbf{h}_1 + \mathbf{b}.
\end{cases}
\end{align}

where the notation follows that of Section \ref{sec:prelim}, and dropout is applied after \(\texttt{Attn}\) and \(\texttt{MLP}\). 
Recall that \(\mathbf{M} \in \mathbbm{R}^{n \times d_{model}}\), and note that \(\mathbf{W} \in \mathbbm{R^{1 \times n}}\). We use a dummy tensor \(\mathbf{q}\) of all \(1s\) to control the dimension of the output and force it to be scalar. Although it is a matrix, we denote \(\mathbf{q} \in \mathbbm{R}^{(1 \times d_{model})}\) in lower-case vector form to make clear that it has only one non-trivial dimension.

The full \(\texttt{step}\) method is given by: 
\begin{equation}
\texttt{step}(\mathbf{X}) \rightarrow \mathbf{X}' =
\begin{cases}
    \mathbf{Z} = \mathbf{X}  \big|_{cat}  \texttt{MLIP}(\mathbf{X}), \\
    \mathbf{H} = \mathbf{W}_2\mathbf{Z} + \mathbf{W}_1\sigma(\mathbf{W}_0\mathbf{Z}), \\
    s = \texttt{dec}(\mathbf{H}), \\
    \mathbf{X}' = s \cdot \texttt{MLIP}(\mathbf{X}) + \mathbf{X}
\end{cases}
\end{equation}

Where \(\mathbf{W}_2 \in \mathbbm{R}^{d_{model} \times 15}\) and if \(d_{model} = 15\) then \(W_2 = I\).

\section{Additional Comparisons}
\subsection{Direct Structure Prediction} \label{sec:is2rs}
Some works in the past have attempted to directly predict the final relaxed structure given an initial structure. We include experiments in Table \ref{tab:is2rs} showing that direct structure prediction methods under perform compared to iterative methods. One notable issue with this is data inefficiency, while a classic MLIP gets \(n\) data points from a trajectory of length \(n\), direct methods get only one. 

\begin{table}[htbp]
    \centering
     \caption{Initial Structure To Relaxed Structure Results} \label{tab:is2rs}
    \begin{tabular}{lr}
        \hline
        \textbf{Dataset} & \textbf{Avg \texttt{Dq}} \\
        \hline
        Si Defects & 14.957\\
        Catalysts &  85.110\\
        Pure Crystals & 87.593\\
        \hline
    \end{tabular}
\end{table}

\subsection{Evaluation with RMSE instead of \texttt{Dq}} \label{sec:rmseLoss}
Theoretical analysis of neural networks tell us that we would expect that any differentiable loss function could be used in the BPTT process \cite{cybenko1989approximation, yun2019transformers, bruel2020universal}. For the core studies of this work, we use \texttt{Dq}, a mass-weighted version of RMSE loss. To demonstrate that this procedure is not limited only to \texttt{Dq} however, experiments using BPTT fine-tuning were conducted on the ADAPT model with RMSE loss on all three datasets. The Si defect dataset model yielded a roughly \(37\%\) decrease in RMSE, the catalysts yielded a roughly \(22\%\) decrease, and the pure crystals leads to numerical instable training. This matches with the observed results in the \texttt{Dq} results, where the high deviation in structures contained in the pure crystals dataset appears to cause somewhat inhibit BPTT training. Instability can likely be accounted for by changing the loss function, as RMSE is known to exhibit unstable training dynamics. 

RMSE is a monotonic transformation of MSE and therefore shares the same minimizers. However, this transformation alters gradient magnitudes, introducing a scaling factor of \( \frac{1}{2 \cdot RMSE} \) which is proportional to the inverse of the error. As a result, gradients can become large near convergence and small during early training, leading to poorer conditioning compared to MSE. Similar sensitivity of optimization dynamics to loss geometry has been observed in prior work on regression losses \cite{qi2020mean}.This could be mitigated through the removal of the square root term, or changing the initialization. Appendix \ref{sec:deepInit} covers more details on mitigating this issue. 

\section{Deep-Network Initialization} \label{sec:deepInit}
Recall that through BPTT training, we are essentially treating the unrolled trajectory as a single neural network. For a \(k\) step relaxation, we are creating a network that is \(k\) times as deep as the original. With \(k\) often in the range of \([50-200]\) in our examples, network depth can become quite extreme. Training a very deep neural architecture is known to increase the likelihood of unstable training. Applying extra normalization steps \cite{ba2016layer, ioffe2015batch, daneshmand2020batch} or changing the network initialization \cite{he2015delving, schoenholz2016deep} has been shown to substantially increase training stability. These offer potential avenues for further development of this work. 

\section{Ablation Studies} \label{sec:ablation}
\subsection{Exploration Noise} \label{sec:noiseInj}
A common technique used in sequence fine-tuning tasks is the addition of noise into model weights or update steps with the intention to force the model to explore weights. Such noise is usually annealed over epochs, as the goal of the model shifts from exploration to exploitation. We conduct studies on the effect of noise in our BPTT approach, and find that it leads to vastly diminished results, even performing worse than the original model before the fine-tuning. This is a surprising result given the findings of \cite{cheon2020crystal} which reports that noise injection improves overall performance.

\subsection{Proximal Term vs Gradient Clipping} \label{sec:proxClip}
Often when performing fine-tuning tasks, a proximal \(\mathcal{L}_2\)-loss term anchoring the model weights to the pretrained version is added to ensure that the model does not stray too far and overfit \cite{li2018explicitinductivebiastransfer}. Other have proposed using aggressive gradient clipping as an alternative \cite{yang2022improvingstabilityfinetuningpretrained}. Our experiments find there is a negligible performance difference between the two approaches. We adopt the gradient clipping becuase it is more computational lighter compared to the proximal-term approach.

\subsection{Use of Momentum and Step Annealing} \label{sec:momentAnneal}
In optimization literature, pure stochastic gradient descent (SGD) is seldom employed in practice. Instead, machine learning typically utilizes step-size schedulers, momentum variants, or second-order approximations. Domain-specific optimizers have also emerged, such as the FIRE algorithm in atomistic modeling. Noise injection---in various forms\footnote{Stochasticity can be seen as a form of ``noise injection'' as well, although we do not test a version of the model without due to compute limitations.}---during training can enhance generalization. Moreover, the use of proximal weight penalties and gradient clipping helps constrain fine-tuning updates, ensuring that new policies remain close to their predecessors and mitigating catastrophic forgetting or overfitting.

Since the relaxation trajectory is inherently an optimization procedure, it is reasonable to expect that such modifications to the PES-level descent may have an impact on the BPTT-derived weight updates. Specifically, we perform ablation studies to test the effect of the PES-level application of: (1) noise injection\footnote{Noise injection has been utilized in literature by \cite{cheon2020crystal}}, (2) momentum and annealing in steps, and (3) intentional varying of trajectory lengths though modifications of stopping conditions. We also test the effect of the MLIP-level usage of proximal term vs gradient clipping update controls. 

\begin{table}[htbp]
    \centering
        \caption{Characterizing the effect of momentum weight of \(0.05\) and \(0.99^k\) annealing-step factor on step \(k\).} \label{tab:momAnnealRate}
    \begin{tabular}{lr}
        \hline
        \textbf{Setup} & \textbf{Avg \texttt{Dq}} \\
        \hline
        ADAPT & 8.47\\
        ADAPT+Momentum & 7.67\\
        ADAPT+Anneal & 7.28\\
        ADAPT+Momentum+Anneal & 7.49\\
        \hline
        \hline
        BPTT & 2.73\\
        BPTT+Momentum & 3.47\\
        BPTT+Anneal & 1928.30\\
        BPTT+Momentum+Anneal & 1969.71\\
        \hline
    \end{tabular}
\end{table}

We find that none of these modifications substantively improve results, and some cause degenerative trajectories. This indicates that: (1) the model is capable of learning a strong set of optimization controls, and (2) these controls are substantively different than those we typically use. We hypothesize that this degradation arises because BPTT primarily improves performance by modifying the MLIP itself to better align with the structural biases and trends exhibited in the dataset, rather than by learning a more effective optimizer \ref{sec:stepSec} \ref{sec:forceStruct}. In our formulation, the relaxation trajectory defines a coupled system between the learned force field and the update rule, and BPTT adapts the MLIP to exploit this specific interaction. Introducing additional optimization heuristics such as momentum or annealing alters the trajectory dynamics, effectively changing the distribution of intermediate states encountered during both training and inference. This can disrupt the learned coupling between the MLIP and the descent process, leading to a mismatch between the optimized behavior and the resulting rollout dynamics. Consequently, these heuristics may interfere with the trajectory-level signal that BPTT leverages, yielding degraded final relaxation performance despite their conventional benefits in standard optimization settings.

Applying a momentum term to the descent term of ADAPT without BPTT yields improved results. Similarly, using a multiplicative annealing---such as \(0.99^k\) on step \(k\)---yields beneficial results. In Table \ref{tab:momAnnealRate}, we report experimental results investigating the effect that these two changes produce when added into the BPTT descent trajectories. We note that the initialization value for the scalar step size hyperparameter is \(0.5\).

\subsection{Intentional Modification of Trajectory Length} \label{sec:intentStop}
During the training process, different examples may run for different numbers of steps. Recalling the view of this problem setup as a deep neural network made of a repeated block, this differing numbers of block instances means that we are optimizing for an inherently different situation in each example. To establish the impact of such potential mismatches, we present experiments in which we artificially vary the stopping threshold after each batch of training. In previous experiments, a threshold of \(0.001\) in total root mean squared error (RMSE) atomic movement across the structure is used, with a maximum of 75 steps being allowed. Under this new paradigm, we randomly varied the threshold between \(0.01\) and \(0.0001\), and randomly varied the maximum number of frames between \(50\) and \(75\). Results from this are presented in Table \ref{tab:stopCond}, and show that such a training setup leads to slightly worse results. The hope of this training scheme--- now disproved---is that it would worsen results on the training set, but could force better generalization and improve test set performance. 

\begin{table}[htbp]
    \centering
     \caption{Characterizing the effect of varying stopping conditions} \label{tab:stopCond}
    \begin{tabular}{lr}
        \hline
        \textbf{Stopping Condition} & \textbf{Avg \texttt{Dq}} \\
        \hline
        Fixed & 2.92\\
        Variable & 3.19 \\
        \hline
    \end{tabular}
\end{table}

\section{Hyperparameters and Reproducibility} \label{sec:hyperparam}
Unless otherwise noted in Section \ref{sec:intentStop} or Section \ref{sec:momentAnneal}, results are from trials using an RMSE threshold of \(0.001\). We impose a maximum of \(75\) frames in a trajectory for Silicon defects and pure crystals, and \(200\) frames for catalysts. The default trajectory-level step size used in Figure \ref{fig:heatmap} is 0.05, although we use \(0.001\) for the PaiNN catalyst example. 

All results were calculated on servers with Nvidia P100 and A40 gpus. 

All data is currently available from the cited resources, and code will be released following paper publication at [public GitHub repository]. 

\subsection{Numerical stability and outlier sensitivity of PaiNN trajectory fine-tuning} \label{sec:painnNotes}
\begin{figure*}
    \centering \includegraphics[width=\textwidth]{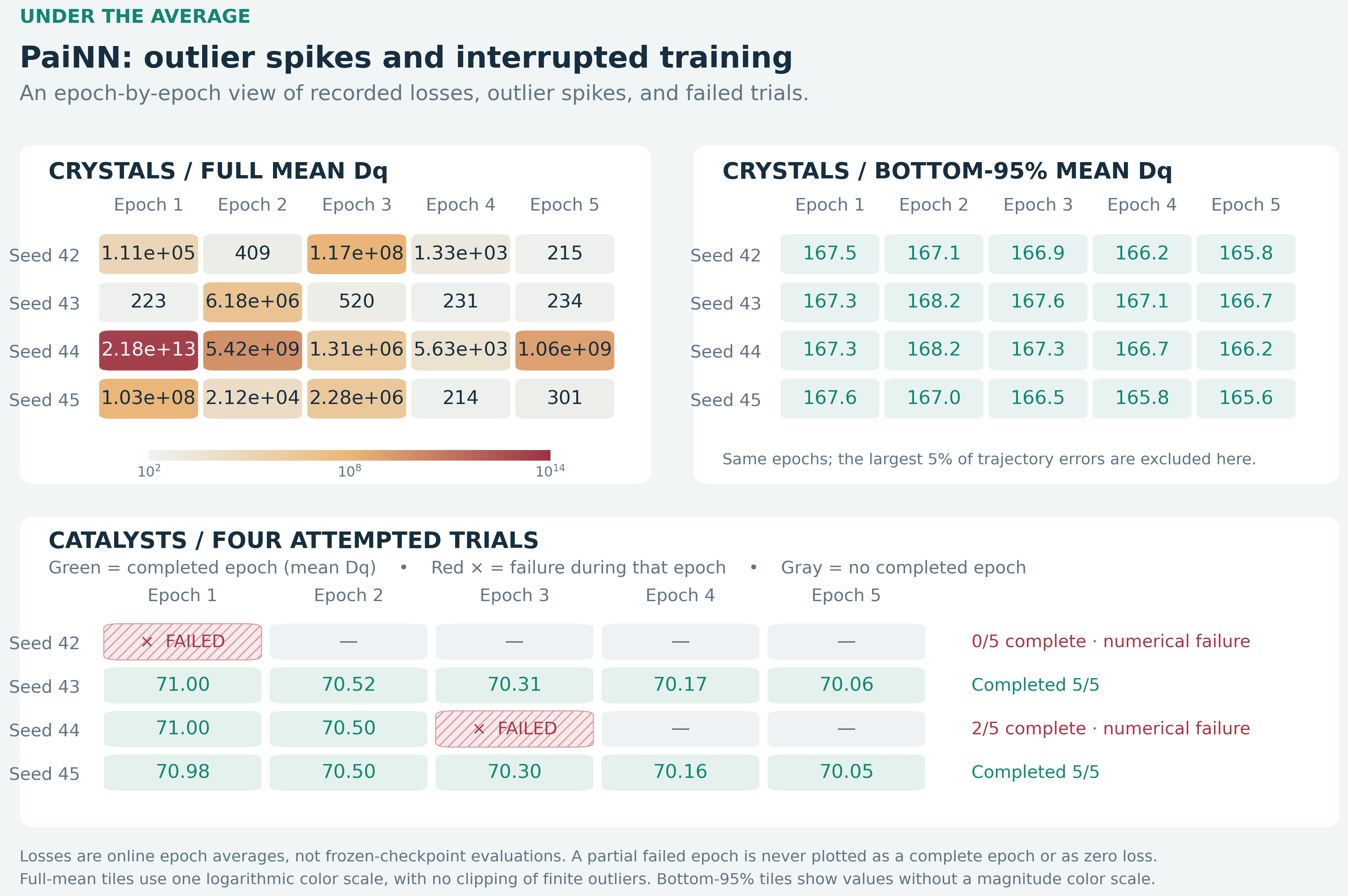} \caption{Results breakdown of numeric instability in PaiNN during BPTT trajectories.} \label{fig:painnStab}
\end{figure*}

Four random-seed trials (42–45) were attempted for each of the three PaiNN datasets, with five epochs of backpropagation through time (BPTT) per trial. Rollouts used nonperiodic geometry, a neighbor list fixed at the initial structure with geometry-dependent quantities recomputed during relaxation, a step size of 0.01, and a displacement-based stopping threshold of 0.001. The maximum rollout length was 100 steps for silicon defects and pure crystals and 200 steps for catalysts. Trajectories were differentiated individually, with gradient accumulation corresponding to a batch size of four. Training structures with more than 4,500 initial directed edges were excluded. Optimization used AdamW with a learning rate of $10^{-5}$ and global gradient-norm clipping at 1.0; gradient norms were evaluated using float64 reductions. These settings define the scope of the following observations.

Ten of the twelve PaiNN trials completed all five training epochs and the final evaluations: four trials each for silicon defects and pure crystals, and two for catalysts. The remaining catalyst trials terminated during training rather than failing a held-out test evaluation. Seed 42 terminated in epoch 1, at batch 1,330, without completing an epoch; seed 44 terminated in epoch 3, at batch 1,084, after completing two epochs. Both recorded the error “Nonfinite accumulated BPTT gradients; weights were not updated.” The failing updates were therefore rejected. These terminal failures were classified as numerical instability, rather than memory exhaustion, in the saved run records (Figure \ref{fig:painnStab}, lower panel).

Accordingly, the reported catalyst endpoint averages include only seeds 43 and 45. Their mean test Dq decreased from 75.247 to 71.955 (4.37\%), whereas test force RMSE increased from 0.24239 to 0.38397 (58.41\%). This comparison supports a structural–force accuracy tradeoff among the completed trials, but it is conditional on completion: two of the four attempted trials failed numerically. It should not be interpreted as a four-trial estimate of performance or as evidence of uniformly reliable training. Failed trials are retained in the completion accounting and are not assigned zero error or included through their incomplete checkpoints. Gradient clipping limits finite update magnitudes, but cannot repair gradients that have already become nonfinite during differentiation. The failure records identify the numerical symptom; they do not establish its underlying physical or architectural cause.

Completion of training did not imply stable crystal rollouts. The full mean Dq was highly sensitive to a small number of extreme trajectories. For example, the online training mean for crystal seed 44 was $2.18 × 10^{13}$ in epoch 1, fell to $5.63 × 10^3$ in epoch 4, and rose again to $1.06 × 10^9$ in epoch 5. Over the same epochs, its bottom-95\% mean Dq remained between 166.24 and 168.18 (Figure \ref{fig:painnStab}, upper panels). Thus, a large decline in the full-mean loss can reflect a reduction in the magnitude of extreme errors without a comparable improvement across most trajectories. The epoch histories are averages accumulated while model parameters change during training; they are distinct from evaluations performed with a fixed checkpoint (Figure \ref{fig:trainingCurves}).

\begin{figure*}
    \centering \includegraphics[width=\linewidth]{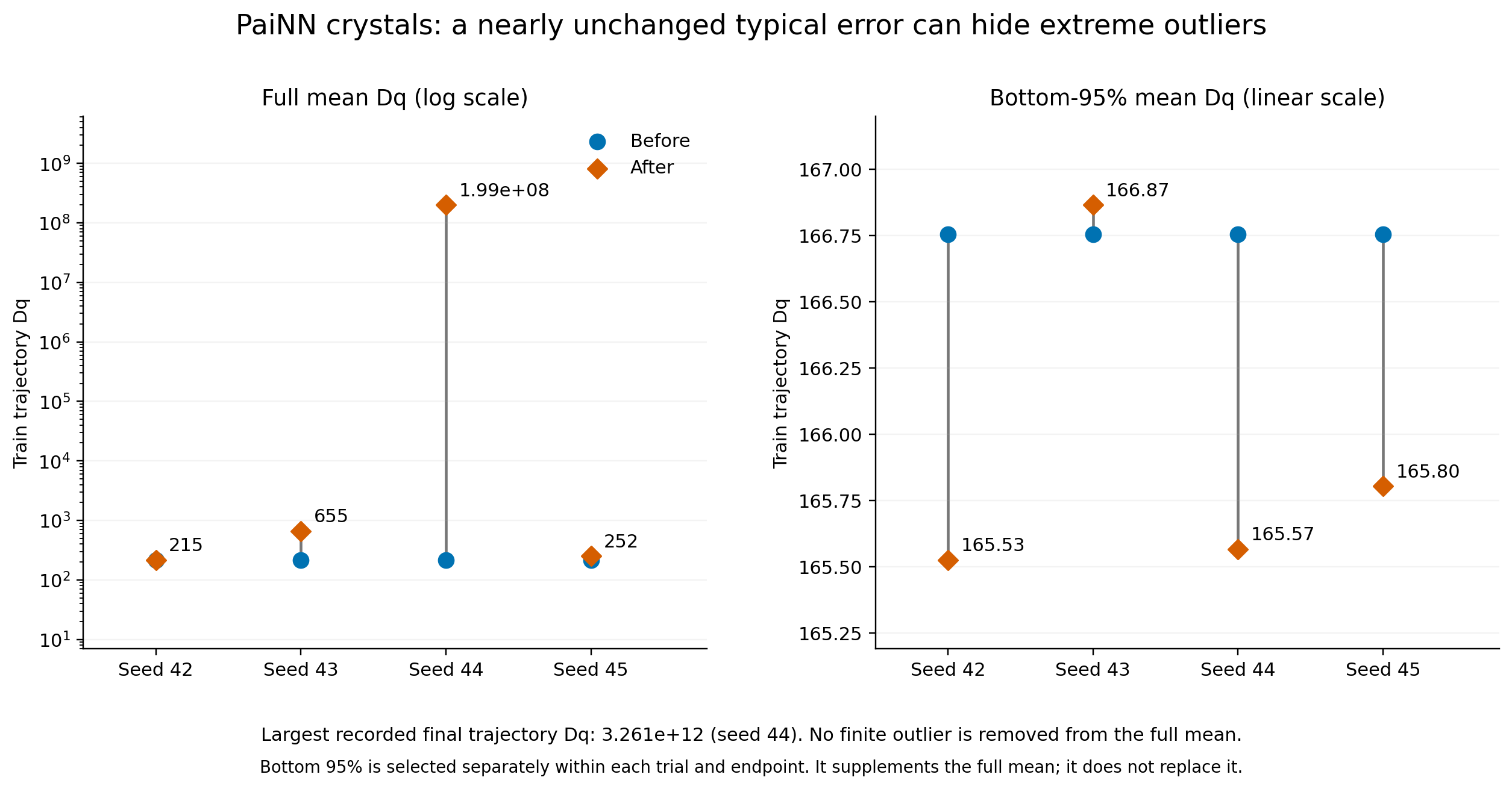} \caption{Demonstrating the effect of the outlier on the averaged statistics. } \label{fig:painnOut}
\end{figure*}

The final fixed-checkpoint evaluation provides a complementary measure of this sensitivity. Across the four crystal seeds, the mean training-set Dq increased from 215.808 before fine-tuning to $4.976 × 10^7$ afterward. The post-training per-seed means were 214.985, 655.205, $1.991 × 10^8$, and 251.694 for seeds 42–45, respectively. In seed 44, one of the 16,382 evaluated training trajectories had Dq = $3.261 × 10^{12}$ and contributed more than 99.999\% of the summed Dq. By contrast, the seed-averaged bottom-95\% training Dq changed only from 166.755 to 165.940, a decrease of 0.49\% (Figure \ref{fig:painnOut}).

Here, the bottom-95\% statistic is the arithmetic mean of the lowest floor(0.95N) trajectory errors within each trial and evaluation. It supplements the full mean rather than replacing it. The retained trajectories are selected separately before and after training and need not be identical. This statistic characterizes performance away from the most extreme errors, but does not establish that individual trajectories improved or that the excluded tail is unimportant. All finite outliers remain included in the full-mean results.

The crystal test results also argue against interpreting the reduction in typical training error as successful generalization. Mean test Dq increased from 119.926 to 122.474 (2.12\%), and bottom-95\% test Dq increased from 100.712 to 104.271 (3.53\%). Both test Dq measures worsened in all four seeds. Test force RMSE increased from 0.36113 to 0.76236 (111.10\%; Figure \ref{fig:heatmap}). The observed outcome is therefore a combination of extreme training-set outliers and modest deterioration in test structural accuracy, rather than a general structural improvement obscured solely by a reporting artifact.

The instability observed in this setting may reflect architectural limitations when differentiating through long relaxation trajectories. Unlike transformer architectures that typically incorporate normalization mechanisms to stabilize deep computation, PaiNN was not explicitly designed for the effective network depth induced by extensive unrolling through BPTT. Nevertheless, this limitation does not invalidate the broader finding. The primary scientific contribution of this work is to demonstrate that optimizing MLIPs for downstream structural relaxation can produce improvements in final-structure accuracy despite substantial degradation in local force fidelity, rather than to establish the universal effectiveness of a particular fine-tuning procedure. Although the PaiNN results indicate that this behavior is not guaranteed across all architectures and datasets, its occurrence in multiple independent experimental settings provides evidence that local force accuracy and downstream relaxation performance need not be aligned.

\section{Datasets} \label{sec:datasets}
Among modern research directions in advanced materials, \textit{crystals}: periodic, infinitely repeating arrangements of atoms known as lattices, are of particular relevance. We validate BPTT fine-tuning by testing on two different datasets which represent fundamentally different types of crystal domains. The success of BPTT fine-tuning under such different domains shows that its benefit is not limited to a specific problem scope.

\subsection{Silicon Crystal Defects} \label{sec:siDefectData}
A central theme in materials design is the deliberate introduction of \textit{defects}: controlled disruptions of an otherwise perfect lattice to engineer specific electrical, magnetic, or quantum properties. This dataset \cite{xiong2023high, xiong2024computationally} focuses on point defects in silicon. Point defects occur wherever the crystal’s regular structure is broken in a constrained local region. They are typically classified into three categories: \(i)\) \textbf{substitution:} a lattice atom is replaced with a different element; \(ii)\) \textbf{interstitial:} an additional atom is inserted between lattice sites; and, \(iii)\) \textbf{vacancy:} a lattice site is missing an atom.

These seemingly simple atomic irregularities, vacancies or irregular atoms within a crystal lattice, are fundamental to modern electronic and quantum technologies. They can also combine to form defect complexes, where multiple point defects interact within the same region of the crystal. Such engineered defects are the basis of many modern technologies such as semiconductor doping for high-performance computing, and they are central to emerging efforts to create stable qubits for quantum communication and sensing applications.

\subsection{CHGNet Pure Crystals Data Subset} \label{sec:CHGNetData}
We use the dataset curation from CHGNet \cite{deng2023chgnet}, which is itself a curation of data from the Materials Project \cite{jain2013commentary}. To avoid prohibitively large runtimes, we select the first \(20,101\) examples as the dataset. This dataset contains pure crystal cells. Each cell is assumed to be subject to the periodic boundary condition, and is part of an infinitely repeating regular lattice. Due to the smaller number of atoms and lack of specific interest areas, we use standard MSE loss as our metric for this dataset. To ensure consistency across testing scenarios and control for potential complicating factors, we train our own base models, first on first predictions, then use BPTT fine-tuning. We show that BPTT leads to a substantial increase in final-structure accuracy, as shown in Figure \ref{fig:chgLoss}. 

Matching to the quantities set in \cite{dramko2025adapt}, we use the first \(100\) trajectories as testing data and the rest as training data. Pure crystal trajectories are often much shorter than those from defects, so we use a substantially larger quantity to create a dataset of similar number of frames to that of the silicon defects from \cite{dramko2025adapt}. The MLIP is first trained on the structure-force pairs from the flattened \(20,001\) training trajectories. After, the model is fine-tuned using BPTT on the \(16,984\) trajectories of length greater than two with constant atom counts throughout the trajectory. Finally, testing results on the \(96\) of the \(100\) test trajectories are reported, four trajectories (sequentially trajectories \(15\), \(17\), \(19\), \(63\)) has mismatched atom counts throughout the trajectory in the dataset, and were thus discarded.

\subsection{Catalyst Dataset}
We use datasets from the open catalyst project \cite{chanussot2021open}. Specifically, the ``s2ef\_train\_200K'' dataset is used to train the force-to-structure underlying model, and the ``is2res\_train\_val\_test\_lmdbs'' is used to train the sequence-level BPTT updates. Matching with the other evaluation setups, the first \(100\) trajectories from the sequence-level dataset are used as the testing case, with the other trajectories providing the training data. As before, we train our own base models to ensure consistency in the testing scenario. 

\section{ADAPT Optimal PES Step} \label{sec:PesStep}
In Table \ref{tab:PESstepParam} we show the average testing \texttt{Dq} for selected values of the PES-level step size. Notably, comparing BPTT against one of these values unfairly disadvantages BPTT because BPTT has no knowledge of the the test set, whereas these hyperparameters are fit directly on the test set. 

\begin{table}
    \centering
    \caption{Hyperparameter Values for PES-Level Descent of ADAPT}
    \label{tab:PESstepParam}
    \begin{tabular}{lr}
        \hline
        \textbf{Step Size} & \textbf{Avg \texttt{Dq}} \\
        \hline
        0.001 & 12.38\\
        0.005 &  9.88\\
        0.007 &  9.15\\
        0.009 & 8.61 \\
        0.011 & 8.23 \\
        0.013 & 7.90 \\
        0.015 & 7.63 \\
        0.017 & 7.46 \\
        0.019 & 7.29 \\
        0.021 & 7.16 \\
        0.025 & 6.90 \\
        0.029 & 6.71 \\
        0.033 & 6.49 \\
        0.037 & 6.35 \\
        0.039 & 6.27 \\
        0.045 & 6.05 \\
        0.049 & 5.91 \\
        0.063 & 5.42 \\
        0.067 & 5.38 \\
        0.071 & 5.33 \\
        0.073 & 5.34 \\
        0.075 & 5.33 \\
        0.077 & 5.32 \\
        0.079 & 5.37 \\
        0.081 & 5.41 \\
        0.083 & 5.40 \\
        0.085 & 5.40 \\
        0.087 & 5.45 \\
        0.091 & 5.35 \\
        0.095 & 5.37 \\
        0.15 & 9.18 \\
        0.25 & 20.46 \\
        \hline
    \end{tabular}
\end{table}

\section{The Effect of Graph Cutoff Building} \label{sec:GraphConstruction}
Both ADAPT and ResMLP utilize coordinate-in-space based architectures. This means that the data representation does not change based on its structural evolution. With graphs however, as the structure evolves, atoms may move in or out of the range of other atoms' cutoff radii. This can result in different representations of the graph over the course of the trajectory. Since this is a non-differentiable decision boundary, we cannot train BPTT directly through graph construction. In many structures, this creates a relatively small problem, as structural evolution often induces in a very limited movement among any given atom in the structure. While BPTT can be implemented to ignore these non-differentiable decisions, it is a potential area of concern for structures with widespread and substantial structural evolution. In our current implementation, we do not consider the PBC during rollout, only during initial model training. It is not impossible to have a PBC-respecting rollout; it is, however, an engineering problem that is orthogonal to our primary goals of showing: (1) that structure information can be used to oversee force information, and (2) that in some domains, reducing force-accuracy can improve relaxation accuracy. 

Additionally, many graph architectures have adopted the implementation choice of using backpropagation to calculate forces from energy predictions. This can result in a very convoluted gradient chain during BPTT, and users must be careful not to consume the computation graph during the force prediction stage. 

\section{Implicit Equilibrium Fine-Tuning with Truncated Trajectory Supervision} \label{sec:gdeq}
Full backpropagation through a relaxation trajectory requires retaining the
computational graph for every force evaluation, causing memory consumption to
grow linearly with the number of relaxation steps. We therefore explored a
hybrid approach that combines a short window of trajectory-level supervision
with ideas from deep equilibrium (DEQ) models to avoid differentiation through
the full relaxation.

\paragraph{Deep equilibrium models.}
DEQ models provide an alternative to explicitly unrolling a repeatedly applied
neural network \cite{bai2019deep}. Consider an iterative transformation
\begin{equation}
    \mathbf{z}_{t+1} = G_\theta(\mathbf{z}_t).
\end{equation}
A conventional recurrent or iterative model evaluates this transformation for
a fixed number of steps and backpropagates through the resulting sequence
\begin{equation}
    \mathbf{z}_0
    \rightarrow
    \mathbf{z}_1
    \rightarrow
    \cdots
    \rightarrow
    \mathbf{z}_T.
\end{equation}
A DEQ model instead treats the output as an equilibrium, or fixed point, of the
transformation. The forward pass solves for a state $\mathbf{z}^{\star}$
satisfying
\begin{equation}
    \boxed{
        \mathbf{z}^{\star} = G_\theta(\mathbf{z}^{\star})
    }.
\end{equation}

Importantly, the intermediate states used to find $\mathbf{z}^{\star}$ need
not be retained for backpropagation. Rather than differentiating through every
iteration, DEQs differentiate the fixed-point equation itself using the
implicit function theorem \cite{bai2019deep}. This produces a gradient involving
\begin{equation}
    \left(
        \mathbf{I} - \mathbf{J}_{G}(\mathbf{z}^{\star})
    \right)^{-1},
\end{equation}
whose action can be computed by solving an associated linear system using
Jacobian--vector or vector--Jacobian products. Consequently, the memory
required for differentiation does not grow with the number of iterations used
to reach equilibrium. A broader literature has subsequently studied both the
stability of these equilibrium models and efficient approximations to their
implicit gradients \cite{bai2021stabilizing, geng2021training}.

Structural relaxation has a natural connection to this formulation. Given a
learned force model $\mathbf{F}_\theta$, we can write one relaxation step as
\begin{equation}
    \mathbf{X}_{t+1}
    =
    G_\theta(\mathbf{X}_t)
    =
    \mathbf{X}_t
    +
    \alpha \mathbf{F}_\theta(\mathbf{X}_t).
\end{equation}
A converged relaxed structure is then a fixed point
\begin{equation}
    \mathbf{X}^{\star}
    =
    G_\theta(\mathbf{X}^{\star}),
\end{equation}
which, for the update above, corresponds to
\begin{equation}
    \mathbf{F}_\theta(\mathbf{X}^{\star}) = \mathbf{0}.
\end{equation}
This suggests that relaxation itself can be treated as a DEQ: run the
relaxation to convergence without retaining its entire computational graph,
and then implicitly differentiate the final equilibrium structure with
respect to the force-model parameters. This differs from recent work applying
DEQs within force-field architectures, where the equilibrium is defined over
internal network features rather than atomic positions \cite{dequify2025}.

However, using only the equilibrium condition discards information about the
trajectory used to reach that equilibrium. This distinction is important in
structural relaxation because the initial structure may lie within a landscape
containing multiple local minima. An endpoint objective can improve the
properties of an equilibrium once it is reached, but provides no explicit
supervision over the intermediate updates that determine which equilibrium is
reached. We therefore combine DEQ-style implicit differentiation with
truncated BPTT, retaining explicit trajectory supervision over the beginning
of the relaxation while treating the final converged state implicitly. This
use of truncated BPTT differs from approaches in which a short unroll is used
primarily as an approximation to the implicit DEQ gradient
itself \cite{geng2021training}; here, the truncated trajectory provides a
separate training signal for the path taken by the relaxation.

\paragraph{Hybrid trajectory and equilibrium supervision.}
Training is divided into three stages. First, we differentiate through only the
first $K$ relaxation steps,
\begin{equation}
    \mathbf{X}_K
    =
    G_\theta^{(K)}(\mathbf{X}_0),
\end{equation}
and apply the trajectory-level structural loss
\begin{equation}
    \mathcal{L}_{\mathrm{guide}}
    =
    \mathcal{L}_{\Delta Q}
    \left(
        \mathbf{X}_K,
        \mathbf{X}_{\mathrm{gt}}
    \right).
\end{equation}
This truncated BPTT stage provides a gradient through the early relaxation
trajectory while requiring memory proportional only to the chosen horizon
$K$. Truncated BPTT is generally a biased approximation to the gradient of a
long recurrent computation \cite{tallec2017unbiasing}; here, however, we use
the truncated gradient intentionally as supervision over the early portion of
the relaxation rather than as a replacement for the final equilibrium
gradient.

Next, $\mathbf{X}_K$ is detached from the computational graph and relaxation
is continued without gradient tracking until a fixed point
$\mathbf{X}^{\star}$ is reached,
\begin{equation}
    \mathbf{X}^{\star}
    =
    G_\theta(\mathbf{X}^{\star}).
\end{equation}
Because the intermediate states of this second stage are not retained, the
relaxation can be continued to convergence without memory increasing with its
length. Standard fixed-point acceleration techniques such as Anderson
acceleration can additionally be used to accelerate this equilibrium
solve~\cite{anderson1965iterative, walker2011anderson}.

Finally, we differentiate the equilibrium loss
\begin{equation}
    \mathcal{L}_{\mathrm{eq}}
    =
    \mathcal{L}_{\Delta Q}
    \left(
        \mathbf{X}^{\star},
        \mathbf{X}_{\mathrm{gt}}
    \right)
\end{equation}
using implicit differentiation. Differentiating the fixed-point equation gives
\begin{equation}
    \nabla_\theta \mathcal{L}_{\mathrm{eq}}
    =
    \frac{\partial \mathcal{L}_{\mathrm{eq}}}
         {\partial \mathbf{X}^{\star}}
    \left(
        \mathbf{I}
        -
        \mathbf{J}_{G}(\mathbf{X}^{\star})
    \right)^{-1}
    \frac{\partial G_\theta(\mathbf{X}^{\star})}
         {\partial \theta}.
\end{equation}
In practice, the inverse is not formed explicitly; instead, its action on the
loss gradient is obtained by solving the corresponding adjoint system using
vector-Jacobian products \cite{bai2019deep}. The combined objective is
\begin{equation}
    \mathcal{J}(\theta)
    =
    \lambda_{\mathrm{guide}}
    \mathcal{L}_{\mathrm{guide}}
    +
    \lambda_{\mathrm{eq}}
    \mathcal{L}_{\mathrm{eq}}.
\end{equation}

Thus, the hybrid method can be viewed as dividing the relaxation into an
explicitly differentiated trajectory and an implicitly differentiated
equilibrium. Only the first $K$ relaxation steps must be stored, reducing the
trajectory-dependent memory requirement from $\mathcal{O}(T)$ for full BPTT
to $\mathcal{O}(K)$, while the remaining relaxation can proceed to convergence
without retaining its computational graph.

\paragraph{Results.}
We evaluated this approach on the same 217-atom silicon point-defect relaxation
task used for trajectory-level supervision. Full BPTT achieved a best test
$\Delta Q$ of $246.0 \pm 5.8$ across five seeds with a peak training memory of
$7.47$~GB. Using the hybrid method with $K=20$ produced a statistically
comparable $\Delta Q$ of $248.3 \pm 9.0$, while reducing peak memory to
$2.15$~GB, a $3.5\times$ reduction, and running $1.82\times$ faster than full
BPTT. Increasing the guidance horizon to $K=40$ gave
$\Delta Q = 251.4 \pm 11.5$ with $4.04$~GB peak memory.

A basin-capture analysis further suggested that the truncated trajectory
component was important rather than the implicit equilibrium objective alone.
The pretrained model reached the correct basin for $43/100$ test structures,
while implicit-only training decreased this to $25/100$. Full BPTT increased
basin capture to $79/100$, and the hybrid approach reached $85/100$. These
results suggest that implicit differentiation can provide a memory-efficient
complement to trajectory-level supervision, but does not replace the need to
differentiate through at least part of the relaxation trajectory. In practice,
the truncation horizon $K$ provides a direct tradeoff between the memory cost
of trajectory supervision and the amount of relaxation history included in
the gradient.

\newpage
\bibliographystyle{apsrev4-2}
\bibliography{bibliography}

\end{document}